\documentclass[a4paper,11pt]{article}
\pdfoutput=1 

\usepackage{jheppub} 
\usepackage{natbib}
\usepackage{times}
\usepackage{float}
\usepackage{amsmath}
\usepackage{graphicx}
\usepackage{amssymb}
\usepackage{slashed}
\usepackage{xcolor}
\usepackage{soul} 
\usepackage{multirow}
\usepackage{bbold}
\usepackage{booktabs}
\usepackage[T1]{fontenc} 
\usepackage[utf8]{inputenc}
\usepackage{braket}

\usepackage{xspace}

\newcommand{\MMHT}{{\rmfamily\scshape MMHT14 NNLO}\xspace}

\newcommand{\be}{\begin{equation}}
\newcommand{\ee}{\end{equation}}
\newcommand{\bea}{\begin{eqnarray}}
\newcommand{\eea}{\end{eqnarray}}
\newcommand{\ba}{\begin{align}}
\newcommand{\ea} {\end{align}}
\newcommand{\SM}{\text{SM}}
\newcommand{\GeV}{\text{GeV}}

\newcommand{\TeV}{~\text{TeV}}

\newcommand{\mBq}{m_{B_q}}
\newcommand{\mDq}{m_{D_q}}
\newcommand{\mDqst}{m_{D^*_q}}

\renewcommand{\Im}{\text{Im}}
\newcommand{\LL}{\mathcal{L}}
\newcommand{\QQ}{\mathcal{Q}}

\newcommand{\Br}{\mathcal{B}}

\def\eq#1{{Eq.~(\ref{#1})}}
\def\eqs#1#2{{Eqs.~(\ref{#1})--(\ref{#2})}}
\def\fig#1{{Fig.~\ref{#1}}}

\def\sec#1{{Sect.~\ref{#1}}}

\def\app#1{{Appendix~\ref{#1}}}

\allowdisplaybreaks

\makeatletter
\g@addto@macro\bfseries{\boldmath}
\makeatother

\title{Exploiting dijet resonance searches for flavor physics}

\author[a,b]{Marzia Bordone,}
\author[c,d]{Admir Greljo,}
\author[e]{and David Marzocca}

\affiliation[a]{Dipartimento di Fisica, Universit\`a di Torino \& INFN, Sezione di Torino, I-10125 Torino, Italy}
\affiliation[b]{Theoretische Physik 1, Naturwissenschaftlich-Technische Fakult\"at, Universit\"at Siegen, Walter-Flex-Stra\ss e~3, D-57068 Siegen, Germany}
\affiliation[c]{Albert Einstein Center for Fundamental Physics, Institut f\"{u}r Theoretische Physik, Universit\"{a}t Bern, \\ Sidlerstrasse 5, CH-3012 Bern, Switzerland.}
\affiliation[d]{CERN, Theoretical Physics Department, CH-1211 Geneva 23, Switzerland}
\affiliation[e]{INFN, Sezione di Trieste, SISSA, Via Bonomea 265, 34136, Trieste, Italy}

\emailAdd{marzia.bordone@to.infn.it}
\emailAdd{admir.greljo@unibe.ch}
\emailAdd{david.marzocca@ts.infn.it}

\preprint{CERN-TH-2021-031, P3H-21-016, SI-HEP-2021-09}

\abstract{In this work, we reinterpret ATLAS and CMS dijet resonance searches to set robust constraints on all hypothetical tree-level scalar and vector mediators with masses up to 5 TeV, assuming a diquark or a quark-antiquark coupling with an arbitrary flavor composition. To illustrate the application of these general results, we quantify the permissible size of new physics in $\bar B_q\to D_q^{(*)+} \,\{\pi, K\}$ consistent with the absence of signal in dijet resonance searches. Along the way, we perform a full SMEFT analysis of the aforementioned non-leptonic $B$ meson decays at leading-order in $\alpha_s$. Our findings uncover a pressing tension between the new physics explanations of recently reported  anomalies in these decays and the dijet resonant searches. The high-$p_T$ constraints are crucial to drain the parameter space consistent with the low-$p_T$ flavor physics data. 
}

\arxivnumber{2103.10332}

\begin{document}
	
\maketitle	
	
\newpage

\section{Introduction}
\label{sec:intro}

Non-leptonic decays are a challenging playground for the Standard Model (SM), particularly for non-perturbative approaches to quantum chromodynamics (QCD). A question which is often neglected in the literature is how much new physics can hide in these decays? In particular, given the present constraints from complementary new physics searches at low and high energies, what is the allowed deviation from the SM predictions? To address this question, we exploit dijet resonance searches at high-$p_T$ colliders as a complementary probe of the hypothetical new physics entering non-leptonic decays. 

We imagine a bosonic mediator $X$ coupled to quarks which, on the one hand, modifies the low-energy meson decays, while on the other hand, can be directly produced and (or) decayed at the LHC by the same interactions. Several challenges are stemming from the lack of knowledge of the underlying microscopic theory beyond the SM. Firstly, there is a broad range of relevant masses that needs to be covered. Here we consider $m_X \in (50, 5000)$~GeV relying crucially on the latest experimental progress on dijet resonance searches by ATLAS and CMS collaborations~\cite{Sirunyan:2018xlo,Aad:2019hjw,Sirunyan:2019vgj,Aaboud:2018fzt}. Secondly, there is a variety of possible representations and flavor couplings of the mediator $X$, for which the expected signal rates in $pp$ collisions differ. One of the main outcomes of our study is the reinterpretation of the existing dijet resonance searches in a general form applicable to flavor physics. Note that, in the narrow-width approximation, the presence of several couplings can only add up in the total $p p \to X \to j j$ rate. This is a very useful feature for flavor physics. Non-leptonic meson decays depend on the product of two couplings when the resonance $X$ is integrated out at tree level. The absence of the signal in the dijet resonance searches sets limits on all $X q^i q^j$ couplings simultaneously, thus also on their product.

In order to exemplify the importance of these results, we investigate  potential new physics (NP) effects in the branching ratio of $\bar{B}_s\to D_s^{(*)+} \pi$ and $\bar{B}\to D^{(*)+} K$ decays. These decays are mediated by the underlying $b\to c \bar{u} d_i$ quark-level transitions, where $d_i = d,s$, rendering their SM theory predictions amongst the most reliable in the sector of non-leptonic decays and are obtained in the framework of QCD factorisation (QCDF) \cite{Beneke:2000ry}.
Since the quarks entering in these decays are distinguishable, topologies like penguin contribution or weak annihilation do not contribute, rendering the description of these decays rather clean. The most up-to-date predictions for the branching ratios have been presented in Ref.~\cite{Bordone:2020gao}. Next-to-leading power corrections, arising at order $\mathcal{O}(\Lambda_\text{QCD}/m_b)$, are found to be subleading compared to the leading-power ones, strengthening the predictive power of QCDF for these channels.

The possibility of NP effects in four-quarks operators has been already entertained, with the focus on low-energy inclusive observables, see for example Refs.~\cite{Bobeth:2014rda,Brod:2014bfa,Jager:2017gal,Chala:2019fdb,Lenz:2019lvd,Cai:2021mlt}. Interestingly, the aforementioned update uncovered an intriguing tension with the data, not yet thoroughly analyzed in the NP context.
A fit to all the available experimental information concerning $\bar{B}_s\to D_s^{(*)+} \pi$ and $\bar{B}\to D^{(*)+} K$ decays is performed, and the current combination of the experimental measurements for their branching ratios is extracted~\cite{Bordone:2020gao}. The comparison with the respective theory predictions shows that the latter always overestimate the former, with a combined discrepancy of about $4.4\sigma$~\cite{Bordone:2020gao}. This trend has already been observed in the literature (see for example Refs.~\cite{Beneke:2000ry,Fleischer:2010ca,Huber:2016xod,Cai:2021mlt}) but has become more apparent due to the updated theory results in Ref.~\cite{Bordone:2020gao}.

A satisfactory explanation of this puzzle is not yet articulated. On the theory side, the hypothesis of a big deviation due to the missing subleading contributions in QCDF seems to be unlikely, since they overshoot the current estimates by at least one order of magnitude. Hence, it seems motivated to entertain the possibility of this deviation being due to NP. In this paper, we try to understand to which extend the NP solution is viable, especially in connection with bounds from related processes, most notably dijet searches at high-$p_T$. The bounds that we obtain from dijet searches can be applied to a broader class of four quark operators, beyond  the ones mediating $b\to c\bar{u}d(s)$ transitions \cite{Grossman:2006jg,Ciuchini:2007hx,Haber:2006ue,An:2009zh,Pirjol:2009vz,Buras:2010zm,Feldmann:2012js,Altmannshofer:2012ur,Ciuchini:2012gd,Franco:2012ck,Engel:2013lsa,Dekens:2014jka,Inoue:2014nva,Assad:2017iib,Bhattacharya:2018msv,Dekens:2018bci,Haba:2018byj,Chala:2019fdb,Altmannshofer:2020shb,Aebischer:2020mkv,Alguero:2020xca}.

The paper is organised as follows. In \sec{sec:highpT} we study dijet searches at the LHC to set generic constraints on hypothetical new resonances. In \sec{sec:highpTpair} we discuss the constraints on the pair production $p p \to X X \to (jj) (jj)$ from gauge interactions and subsequent decay into jets, while in \sec{sec:highpTsingle} we discuss the single dijet resonance production $p p \to X \to jj$. In \sec{sec:EFT} we present a complete leading-order effective field theory (EFT) analysis of $b\to c \bar{u} d_i$ transitions. In \sec{sec:mediators_dijet} we consider all simplified tree-level mediator models able to explain the aforementioned anomaly and confront the relevant parameter space against the high-$p_T$ collider constraints. The $\Delta F = 2$ transitions are typically induced alongside and require a specific structure to comply with the present bounds while giving sufficient contribution to non-leptonic decays. For two explicit models, color-sextet scalar and colorless weak-doublet scalar, we perform a thorough study of the flavor phenomenology to find the parameter space consistent with the stringent bounds from $\Delta F = 2$ and other observables, while still being capable to address the anomaly. Nevertheless, we show that the remaining parameter space is (almost) entirely ruled out by the dijet searches from \sec{sec:highpT} pushing the models towards a strongly coupled regime featuring broad resonances beyond the validity of perturbation theory. This serves as a practical example of complementarity between low-energy flavor physics and the high-$p_T$ searches at the LHC. Finally, many details of the calculations are left for Appendices, and we conclude in \sec{sec:Conc}.

\section{Dijet searches at high-$p_T$}
\label{sec:highpT}

\begin{figure}[t]
\centering
\includegraphics[height=7cm]{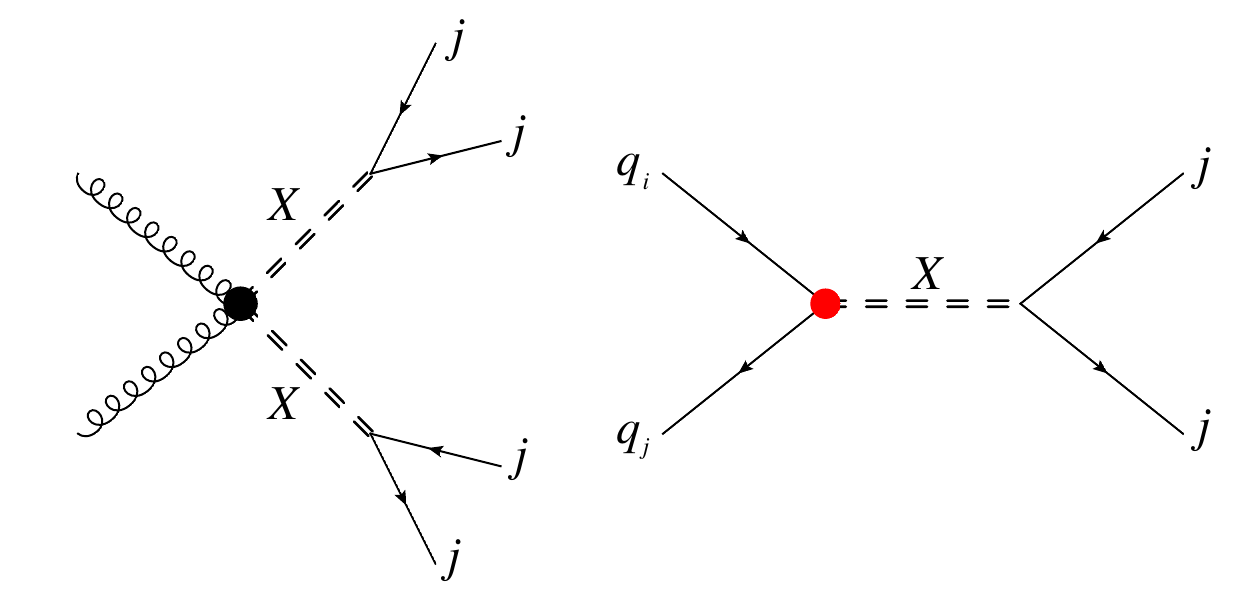}
\caption{\label{fig:Xprod} Representative Feynman diagrams for the pair production $p p \to X X \to (jj)(jj)$ ({\bf left diagram}) and the single production $p p  \to X \to jj$ ({\bf right diagram}) of a dijet resonance $X$ at the LHC. The constraints from the existing searches are reported in Section~\ref{sec:highpT} for different representations and flavor interactions.}
\end{figure}

In this Section we reinterpret the latest ATLAS and CMS dijet searches in terms of constraints on a generic resonance $X$ coupled to a pair of SM quarks of an arbitrary flavor, decaying predominantly into jets. There are two inevitable production mechanisms of $X$ at high-energy hadron colliders. First, when $X$ is charged under a nontrivial SM gauge representation it can be pair produced via gauge interactions.
Second, the dijet resonance can be singly produced directly from quark collisions. The representative diagrams of the two production mechanisms are shown in Fig.~\ref{fig:Xprod}. The left one corresponds to QCD pair production, in the case of a colored resonance, which is fairly large at hadron colliders {(other diagrams are not shown for simplicity)}, while the diagram on the right represents the single dijet resonance production. In the most general case, when additional (sizeable) interactions are present, the resonance decays (promptly) to either dijet, charged leptons, top quark, electroweak gauge bosons, or exotic charged particles. For comparable rates, the dijet final state is hardest to detect at hadron colliders due to the overwhelming QCD background. The constraints in this section are obtained assuming $\mathcal{B}(X  \to  jj) = 1$. While the rescaling for different $\mathcal{B}$ is straightforward, if other decay channels are present it might be worth considering the constraints from the corresponding searches since they may be stronger than those from dijets.

The total decay width to mass ratio $\Gamma_X / m_X$ is a crucial parameter in resonance searches. We will focus mostly on models featuring narrow resonances in which perturbative calculations are fully under control. By the optical theorem, the total decay width is related to loop corrections to the propagator. Collider searches for narrow resonances typically imply the following condition, $\Gamma_X / m_X \lesssim 0.1$, due to the limited detector resolution. In the second part of the study, where we focus on the NP explanation of $b \to c \bar u d_i$ anomalies, the narrow width approximation is valid in a broad mass range assuming the minimal set of couplings. Nonetheless, we will comment on how much the collider bounds can be relaxed for a broad resonance, with an increased $\Gamma_X / m_X$ ratio, while being cautious about the validity of the calculation.

The resonances from new dynamics may or may not be within the kinematical reach of the LHC. If the resonance mass $m_X$ is above the reach for on-shell production at the LHC, its effect can be studied in the high-$p_T$ dijet tails in terms of  four-quark contact interactions. From the experimental point of view, this requires a qualitatively different approach since it is no more possible to fit the data with a resonance-like signal over a smooth background. Other observables, such as the angular distributions of the two jets, are instead employed. For example, see Ref.~\cite{Aaboud:2017yvp} for an ATLAS search and Ref.~\cite{Alioli:2017jdo} for and EFT analysis in terms of flavor-universal contact interactions. We leave the analysis of the full set of flavor-dependent four-quark contact interactions for future work, focusing here on on-shell narrow resonances. The non-leptonic decays studied in the second part of the paper focus on the weakly coupled ultraviolet (UV) completions for which the resonance searches are sufficient, while contact interactions will be relevant for strongly coupled UV completions.

\subsection{Pair production of dijet resonances}
\label{sec:highpTpair}

Even when the couplings to quarks are small, the resonance $X$ is pair-produced by gauge interactions, as in the left diagram of Fig.~\ref{fig:Xprod}. The pair production rate is robustly set by the resonance mass $m_X$ and its gauge representation. We further assume $X$ undergoes a prompt decay to a dijet final state. The LEP-II bounds rely on QED production in $e^+ e^- \to X \bar X$ and apply for all electrically charged resonances. A narrow scalar resonance exclusively decaying to $jj$ is ruled out, unless
\begin{equation}
{\rm (LEP-II)} \; \; \; m_{X^{\pm 1/3}} \gtrsim 80\,{\rm GeV}\,,~m_{X^{\pm 1}} \gtrsim 95\,{\rm GeV}~,
\end{equation}
see~Fig.~9~(c)~in~\cite{Heister:2002jc}. Similar limits apply for vector resonances.

Tevatron and LHC bounds require QCD interactions to be effective and thus apply only to colored resonances. QCD pair production of colored resonances at hadron colliders is overwhelming. The main challenge in these searches is to suppress the large multijet background. Nonetheless, the most recent ATLAS and CMS searches at 13~TeV with about $36\,$fb$^{-1}$ can robustly exclude pair-produced colored resonances decaying exclusively to $jj$~\cite{Aaboud:2017nmi,Sirunyan:2018rlj}. In particular, the experimental limits on the complex scalars, for different color and weak representations under $(SU(3)_c,SU(2)_L)$,  are
\begin{center}
\begin{tabular}{ |c|c|c| c| } 
 \hline
 Scalar & ({\bf 3},{\bf 1}) & ({\bf 6},{\bf 1}) & ({\bf 8},{\bf 1})\\ 
  \hline
\multirow{2}{2.5em}{$m_X > $} &  $410$\,{\rm GeV}~(ATLAS) &  820\,{\rm GeV}~(ATLAS) & $1050$\,{\rm GeV}~(ATLAS) \\ 
  & $520$\,{\rm GeV}~(CMS)& 950\,{\rm GeV}~(CMS) & 1000\,{\rm GeV}~(CMS)\\
 \hline
  Scalar & ({\bf 3},{\bf 3}) & ({\bf 6},{\bf 3}) & ({\bf 8},{\bf 2})\\ 
  \hline
\multirow{2}{2.5em}{$m_X > $} &  $620$\,{\rm GeV}~(ATLAS) &  1200\,{\rm GeV}~(ATLAS) & $1200$\,{\rm GeV}~(ATLAS) \\ 
  & $750$\,{\rm GeV}~(CMS)& 1200\,{\rm GeV}~(CMS) & 1200\,{\rm GeV}~(CMS)\\
 \hline
\end{tabular}~.
\end{center}
Here we report the upper edge of the exclusion mass window, while the lower edge extends down to the LEP-II exclusions. In other words, the combination of all experiments robustly excludes a resonance $X$ with the mass smaller from what is reported in the table above. The limits on the color triplet and octet are directly based on the stop and sgluon benchmarks, respectively. Note that {the color} octet is a complex field, which doubles the sgluon cross section used in~\cite{Aaboud:2017nmi}. We neglect small differences in the acceptance times efficiency for resonances of a different color (and spin), such that representations not considered by the experimental collaborations are constrained by comparing the predicted production cross sections with the 95\% confidence level (CL) observed limits from Figure 9 of Ref.~\cite{Aaboud:2017nmi} and Figure 11 of Ref.~\cite{Sirunyan:2018rlj}. This is validated comparing the exclusion limits on stop, sgluon, and coloron from Ref.~\cite{Aaboud:2017nmi}. For the color sextet, we calculate the cross section using {\tt MadGraph5\_aMC@NLO}~\cite{Alwall:2014hca} and the {\tt UFO} model from the {\tt FeynRules}~\cite{Alloul:2013bka} repository based on the implementation of~\cite{Han:2009ya}.

The limits on the vector resonances depend on the UV completion. For example, a vector color triplet can have an additional non-minimal coupling of the type $\mathcal{L} \supset - i g_s \kappa \,  X^\dagger_\mu T^a X_\nu G^{a\mu\nu}$. If the resonance is a massive gauge boson left after the breaking of extended gauge symmetry, the Yang-Mills {(YM)} case $\kappa=1$ applies. Another example is the minimal coupling {(MC)} case $\kappa = 0$, which usually leads to conservative colliders constraints. Reinterpreting the searches~\cite{Aaboud:2017nmi,Sirunyan:2018rlj} for these two cases, we find
\begin{center}
\begin{tabular}{ |c|c|c| } 
 \hline
Vector ({\bf 3})  & YM & MC \\ 
  \hline
\multirow{2}{2.5em}{$m_X > $} &  $1150$\,{\rm GeV}~(ATLAS) &  700\,{\rm GeV}~(ATLAS) \\ 
  & $1150$\,{\rm GeV}~(CMS)& 800\,{\rm GeV}~(CMS) \\
 \hline
\end{tabular}~.
\end{center}
The appropriate cross sections are calculated using the {\tt UFO} model from Ref.~\cite{Dorsner:2018ynv}. As shown in this example, the limits on vector resonances are extremely sensitive to the UV completion. Nonetheless, the exclusions are typically stronger compared to their scalar counterparts. 

As a final comment, when a resonance has sizeable couplings to valence quarks, there is an additional contribution to production $q \bar q \to X \bar X$ with $t$-channel quark exchange. Since the overall rate is dominated by the gluon fusion, the (potential) negative interference with the sub-dominant $q \bar q$ diagram has no practical impact on the limits. On the other hand, when the coupling gets larger, the total cross section is increased, and the previously quoted exclusions become even more stringent. (For a related study see Fig.~3 in Ref.~\cite{Dorsner:2014axa}.) However, given the limits on the couplings derived from the single production discussed below, the $t$-channel contribution to the pair production can be safely neglected.

\subsection{Dijet resonance}
\label{sec:highpTsingle}

The coupling of the field $X$ to an arbitrary pair of quarks necessarily leads to a resonant production of $X$ in $pp$ collisions at high enough energies, followed by a dijet decay signature as shown in Fig.~\ref{fig:Xprod}~(right). Experiments have searched for a new dijet resonance, and set competitive constraints over a wide range of $m_X$.
We can use these null results to set robust upper limits on the size of {the $X$ coupling to any quark pair, as a function of $m_X$.} The idea is the following --- the different flavor channels $q^i q'^j \to X$ add up incoherently in the total cross section --- thus an upper limit on $p p \to X \to j j$ simultaneously bounds the absolute values of all $X q^i q'^j$ couplings. We carry out a general analysis of the latest ATLAS and CMS dijet searches~\cite{Sirunyan:2018xlo,Aad:2019hjw,Sirunyan:2019vgj,Aaboud:2018fzt} in the mass range \mbox{$m_{W'} \in (450, 5000)$~GeV}  for all possible spin-zero and spin-one mediators considering the most general flavor structure for the couplings.

\subsubsection*{$W'$ example}

To set up the stage, let us consider a benchmark example. The partial decay width for a spin-one colorless $W'$ resonance with the interaction Lagrangian
\be
\mathcal{L} \supset x_{i j}~ \bar u^i_L \gamma^\mu d^j_L ~ W'_\mu + \textrm{h.c.}~,
\ee
is given by
\be
\Gamma_{W' \to u^i \bar d^j} = \frac{m_{W'}}{8 \pi} |x_{ij}|^2~.\label{eq:decGM}
\ee
The leading-order cross section for the production of a narrow positively-charged resonance $W'$ in the quark fusion ($u_i \bar d_j \to W'$) at the LHC is determined by the partial decay width of the inverse process, 
\begin{equation}
\sigma(p p \to W') = \frac{8  \pi^2}{3 s_0} ~ \frac{\Gamma_{W' \to u^i \bar d^j}}{m_{W'}}~\int_{\tau}^1 dx\,~\frac{1}{x} \, f^p_{u^i}(x) \, f^p_{\bar d^j}(\tau /x)~,\label{eq:xsec}
\end{equation}
where $\tau=m^2_{W'}/s_0$ with $\sqrt{s_0}$ the collider energy, and $f^p_q(x)$ are the parton distribution functions evaluated at the factorisation scale $\mu_F = m_{W'}$. In the considered mass range the higher-order radiative corrections of the inclusive production cross section are expected to be of $\mathcal{O}(10\%)$.
In our numerical calculations we use \MMHT central PDF set~\cite{Harland-Lang:2014zoa}. Analogous expressions are for the charged-conjugate process. 

Upper limits on the coupling as a function of the mass are extracted from the ATLAS and CMS exclusions on a specific flavor-universal $Z'$ benchmarks~\cite{Aaboud:2018fzt,Sirunyan:2018xlo,Aad:2019hjw,Sirunyan:2019vgj}. These analyses hunt for a resonance in the invariant mass of the two highest-$p_T$ jets consistent with the inclusive $ p p \to Z' \to j j $ kinematics.\footnote{Instead, dedicated searches for the low-mass resonances target $ p p \to Z' j \to (j j) j$ process~\cite{Sirunyan:2019vxa,Sirunyan:2019pnb}. These probe the parameter space untouchable by previous experiments at lower energies (see e.g. CDF~\cite{Aaltonen:2008dn} and UA2~\cite{Alitti:1993pn}). However, the presence of the additional jet complicates the reinterpretation of the bounds for other spin and flavor cases due to the huge QCD corrections from $g q$ and $gg$ induced diagrams~\cite{Rubin:2010xp}. In Section~\ref{sec:Higgs} we recast~\cite{Sirunyan:2019vxa} to set limits on a $H'$ model example.} The coupling versus mass limits are {adapted} to our cases by equating the total production cross sections for two models in the fiducial region. Fig.~\ref{fig:dijet} (top panel) shows exclusions on a $W'$ coupled to first-generation quarks, $x_{u d} \neq 0$. The plot shows 95\% CL upper limits on $|x_{u d}|$ in the mass range $m_{W'} \in (450, 5000)$~GeV. The vertical axis on the right shows the corresponding partial decay width $\Gamma_{W'} / m_{W'}$ in \eq{eq:decGM}, justifying the narrow-width approximation.

\subsubsection*{$X u d$ couplings}

Let us now reinterpret these bounds for a bosonic complex resonance $X$ (scalar or vector) of any color representation $r \in ({\bf 1}, {\bf 3}, {\bf 6}, {\bf 8})$.
We define the coupling of $X$ to up and down quarks of arbitrary flavors $i$ and $j$ as $x_{ij}$ in the interaction Lagrangians in Eq.~\eqref{eq:X_lagrangians}.\footnote{We do not consider non-renormalisable derivative interactions.}

\begin{table}[t]
\centering
\quad \begin{tabular}{|c|c|c|c|c|}
\hline
$SU(3)_c$ & \bf{1} & \bf{3} & \bf{6} & \bf{8} \\ \hline
$\delta_C$ &  1 & 2 & 2 & 4/3\\
\hline
$\gamma_C$ &  1 & 2/3 & 1/3 & 1/6 \\ \hline
\end{tabular}
\quad
\begin{tabular}{|c|c|c|}
\hline
spin & \bf{0} & \bf{1}  \\ \hline
$\delta_S$ & 1/2 & 1 \\ \hline
$\gamma_S$ & 3/2 & 1 \\ \hline
\end{tabular} \\
\caption{Production cross sections ($\delta$) and decay widths ($\gamma$) rescaling factors for resonances $X$ of  different color and spin with $Xud$ couplings. For more details see \sec{sec:highpTsingle} and Fig.~\ref{fig:dijet}. The interaction Lagrangians are defined in \app{sec:mediators}.}
\label{tab:factors}
\end{table}

\begin{figure}[t]
\centering
\includegraphics[height=8.5cm]{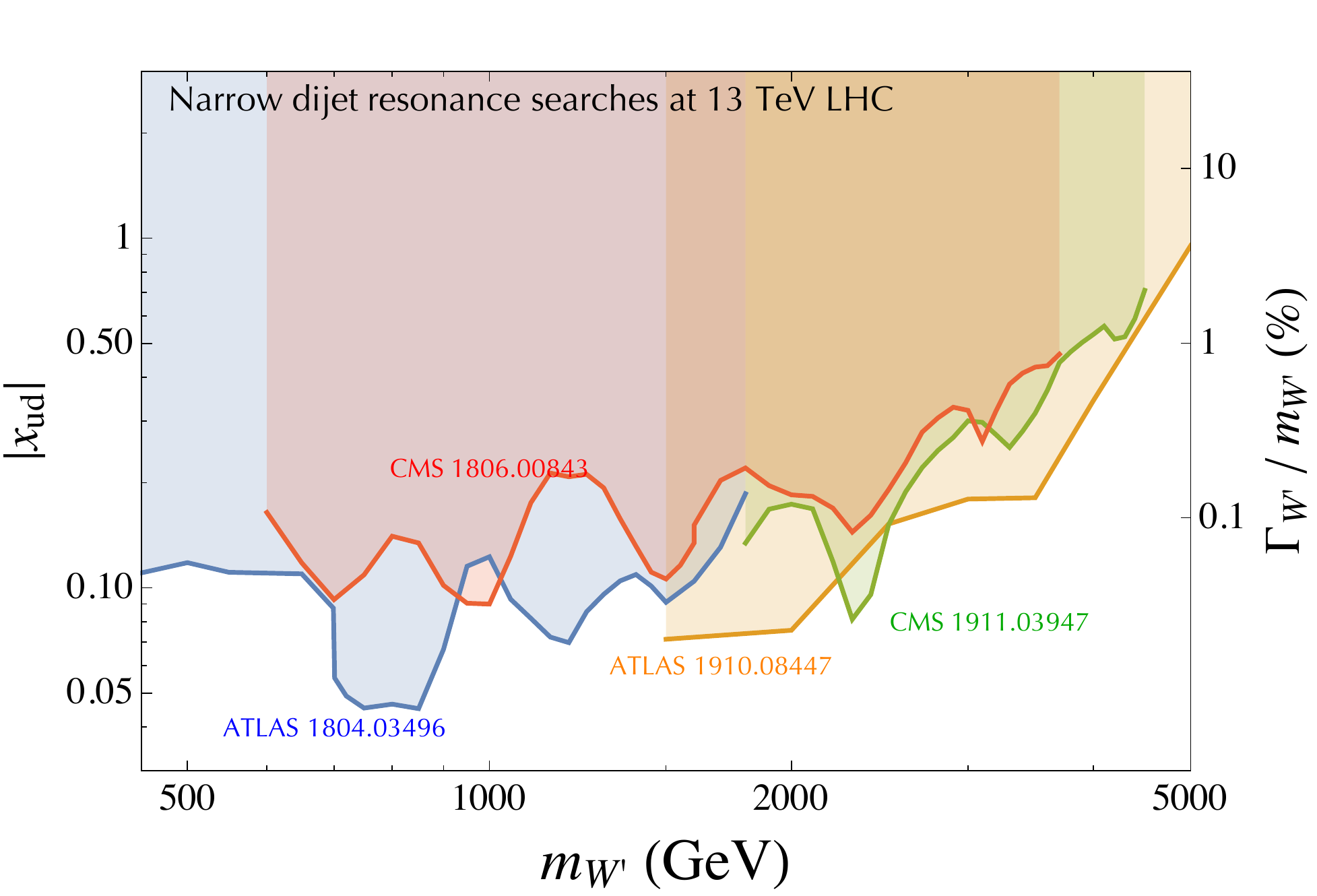}
\includegraphics[height=8.5cm]{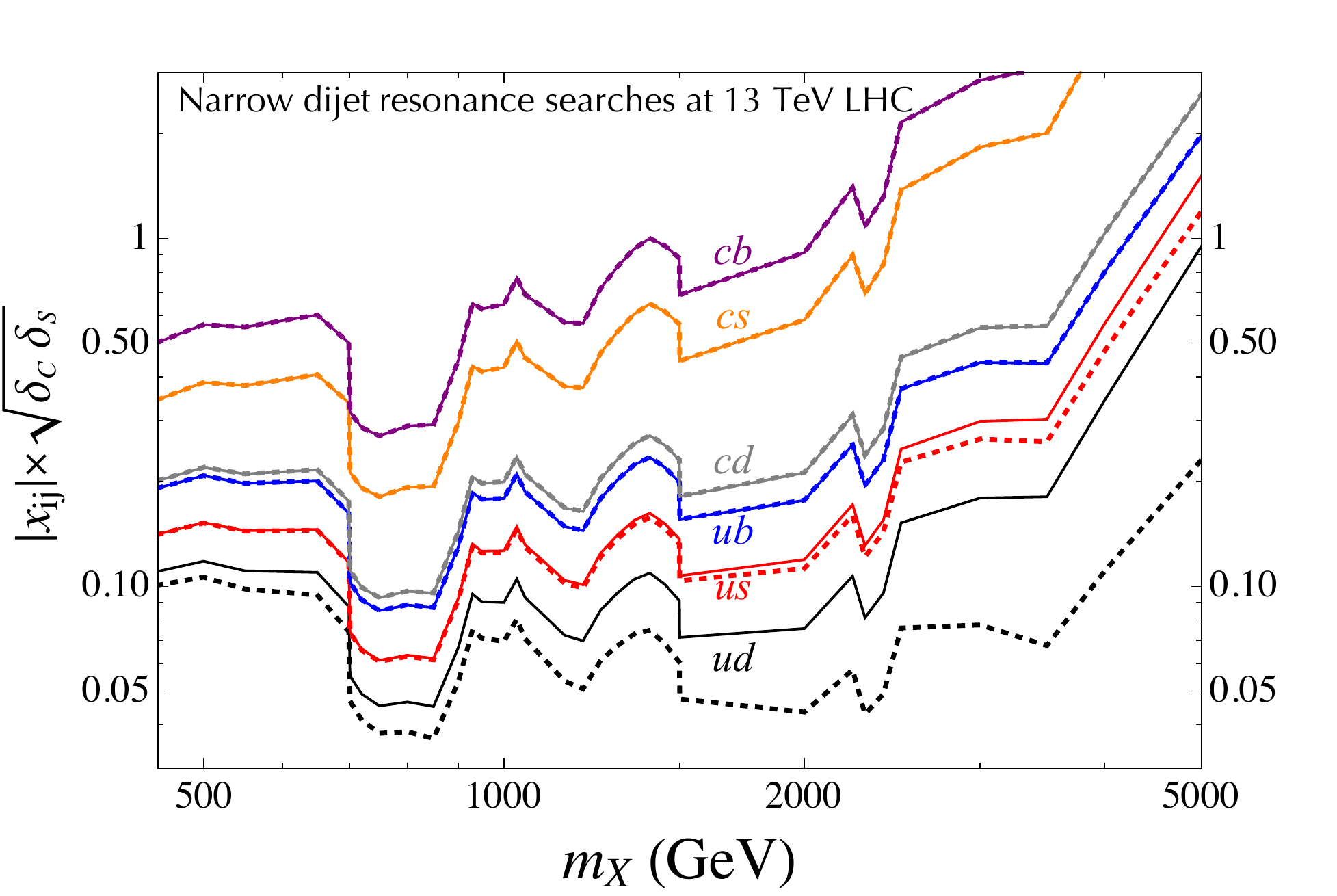} 
\caption{\label{fig:dijet} Experimental limits on a narrow dijet resonance from the LHC searches at 13 TeV~\cite{Aaboud:2018fzt,Sirunyan:2018xlo,Aad:2019hjw,Sirunyan:2019vgj}. {\bf Top panel} is for a spin-1 colorless $W'$ coupled to a single flavor combination of chiral quarks, $x_{ud}\neq0$. The plot shows upper limits at 95\% CL on the absolute value of the coupling from several CMS and ATLAS searches. The vertical axis on the right-hand side is the corresponding partial decay width $\Gamma_{W'} / m_{W'}$ from \eq{eq:decGM}.  {\bf Bottom panel} shows the combined dijet limits on resonances of different spin and color, as well as, arbitrary flavor couplings $ij$. Dashed lines are for diquark resonances (color triplets and sextets) while solid lines are for quark-antiquark resonances (color singlets and octets). The multiplicative rescaling factors for color ($\delta_C$) and spin ($\delta_S$) are reported in Table~\ref{tab:factors}. This plot assumes $\mathcal{B}(X\to jj)=1$ and is valid when the total decay width to mass ratio is $\Gamma_X / m_X \lesssim 10\%$. }
\end{figure}

Dijet searches do not discriminate well between different diquark and quark-antiquark resonances, see the discussion in Ref.~\cite{Sirunyan:2018xlo}. In other words, the $W'$ results can be reinterpreted for other mediators $X$ with different color, spin, and flavor couplings. Comparing the production cross sections for the same coupling $x_{i j}$, we find
\be
\frac{\sigma(p p \to X)}{\sigma(p p \to W')} = \delta_C \delta_S~,\label{eq:Xfactor}
\ee
where the color and spin $\delta$ factors are reported in Table~\ref{tab:factors}. For convenience, we also report the color and spin $\gamma$ factors for resonances $X$, defined as
\be
\frac{\Gamma_{X \to u^i d^j}}{m_X} = \gamma_C \gamma_S \, \frac{\Gamma_{W' \to u^i \bar d^j}}{m_{W'}}  ~.
\ee 
The exclusions on the couplings for other cases are obtained by rescaling those in \fig{fig:dijet} (top panel) with the appropriate ratio of parton luminosity functions.
The generic constraints are shown in \fig{fig:dijet} (bottom panel) when combining all experimental searches. Dashed lines are for $q q \to X$ while solid lines are for $q \bar q \to X$. The two are different only for the  $u d$ versus $u \bar d$ case as expected for valence quarks. The first set should be used for color triplet and sextet resonances, while the second for color singlets and octets.

\begin{figure}[t]
\centering
\includegraphics[height=8.5cm]{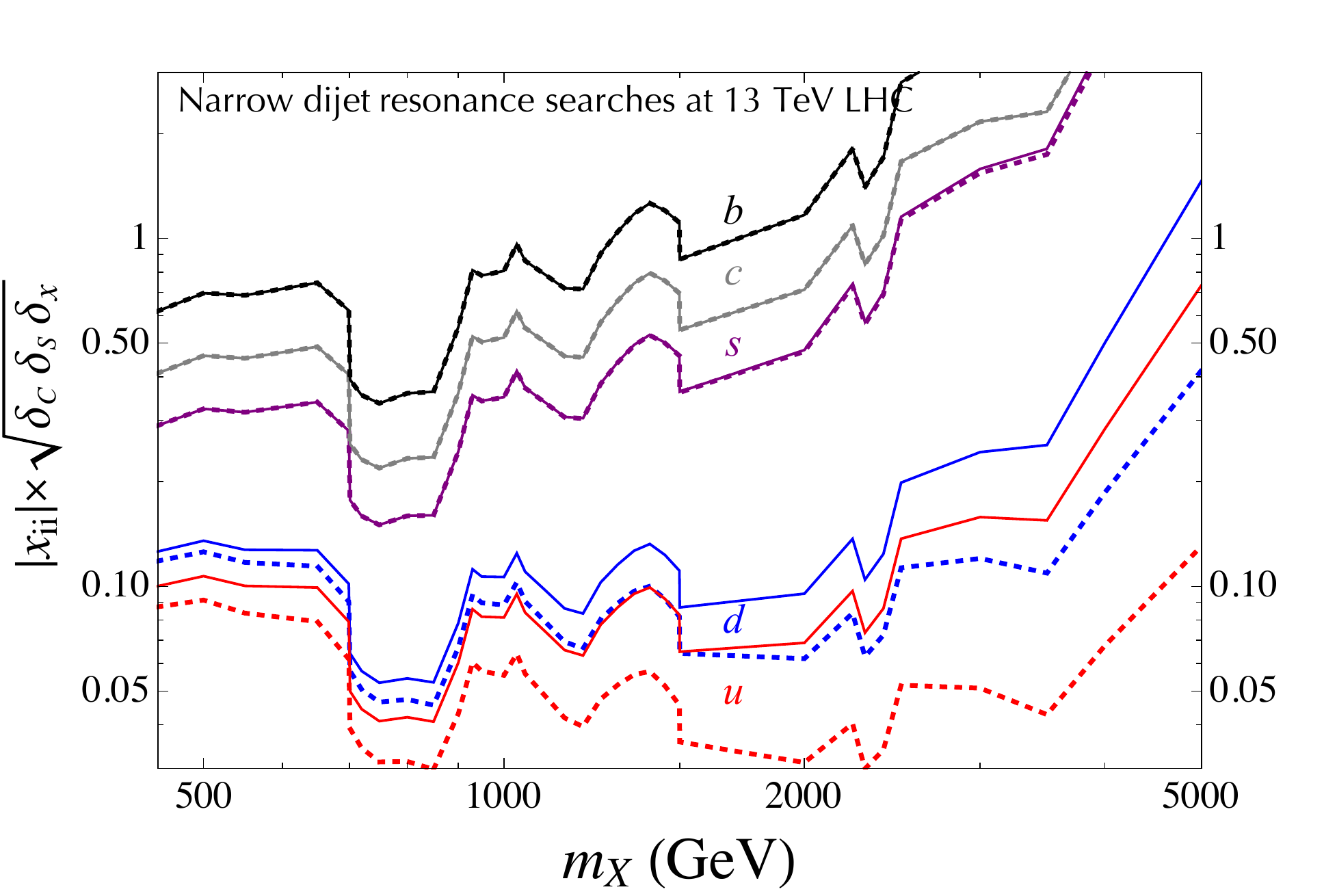}
\includegraphics[height=8.5cm]{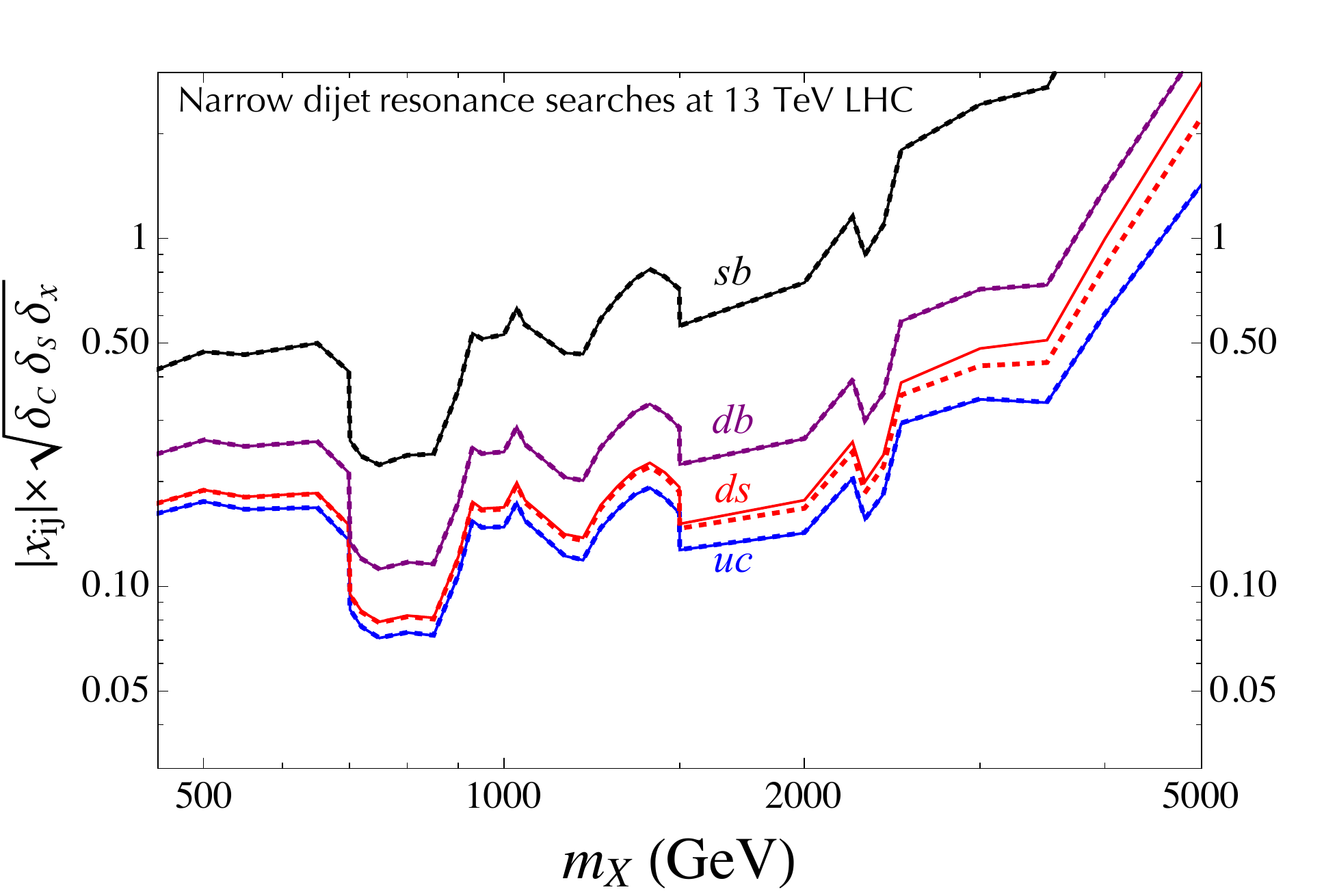} 
\caption{\label{fig:dijetNC} Same as Fig.~\ref{fig:dijet} but for a pair of up or a pair of down quarks. The top figure is for the flavor diagonal couplings while the bottom one is for the flavor changing. For more details see \sec{sec:highpTsingle}.}
\end{figure}

\subsubsection*{$X u u$ and $X d d$  couplings}

Also in this case we define the interaction Lagrangian in Eq.~\eqref{eq:X_lagrangians}, assuming both quarks are either up-type or down-type. We describe electrically neutral resonances with real fields, see the footnote below Eq.~\eqref{eq:X_lagrangians}.
The constraints are summarized in Fig.~\ref{fig:dijetNC} for the flavor diagonal couplings (top plot), as well as, for the flavor violating (bottom plot). The rescaling factors $\delta_C$ and $\delta_S$ are the same as in Table~\ref{tab:factors}. The $\delta_x$ factor equals 1 with the exception of the diagonal couplings for spin-1 color singlet and octet $\delta_x= 1/2$, the off-diagonal couplings for spin-0 color triplet and sextet $\delta_x= 4$, and the diagonal couplings for spin-0 color sextet $\delta_x = 2$.

\subsubsection*{Discussion}

We study the variance of the $p p \to X \to j j$ kinematics for different spins and flavor couplings and their impact on the reinterpretation of the ATLAS and CMS bounds derived in a specific model. To this purpose we use {\tt MadGraph5\_aMC@NLO}~\cite{Alwall:2014hca} to simulate the process at the partonic level. Dedicated {\tt UFO} models for vector and scalar resonances are obtained with {\tt FeynRules}~\cite{Alloul:2013bka}. Motivated by the signal acceptance criteria in Ref.~\cite{Aaboud:2018fzt}, we impose the following partonic-level cuts $p_T(j_1) > 200$\;GeV, $p_T(j_2) > 100$\;GeV, and $|\eta_{1,2}| < 2.8$. We also require the pseudorapidity difference $| \eta_1 - \eta_2| < 1.2$ and the dijet invariant mass $m(jj)> 450$\,GeV. We study the benchmark point $m_X=0.75$\,TeV and $\Gamma_X/m_X = 0.06$. We find that the cut efficiency for the scalar is about $15$\% larger than for the vector. This comes from the pseudorapidity cuts, in particular, for the scalar, the $2 \to 2$ differential partonic cross section $d\sigma / d t$ does not depend on the Mandelstam variables $t$ and $u$, while for the vector, it is proportional to $t^2$ or $u^2$. Thus, our bounds on the scalar resonances are somewhat conservative since the collaborations model the signal with the vector resonance. The above exercise was performed for the coupling with the first generation only. We repeat the simulation for the scalar resonance for all flavor combinations of incoming partons: $u d$, $u s$, $u b$, $c d$, $c s$, and $c b$. The difference in the signal acceptance is at most a few percent between different flavors.

Theoretical uncertainties on the production cross section are due to higher-order radiative corrections, as well as limited knowledge of parton luminosity functions. The most important NLO QCD corrections depend on the mediator representation and typically increase the rate by~$\mathcal{O}(10\%)$, see example Ref.~\cite{Han:2009ya} for the color sextet scalar. In this respect, our leading-order calculation gives somewhat conservative bounds on the coupling. Note that the relative error on the upper bound on $|x_{i j}|$ is half of the relative error on the cross section.

The other source of uncertainties comes from the determination of the parton luminosity functions. Relative uncertainties on the parton luminosities (the integral in \eq{eq:xsec}) are shown in Fig.~5.10 of Ref.~\cite{Ball:2017nwa} for three different PDF collaborations: NNPDF3.1~\cite{Ball:2017nwa}, CT14~\cite{Dulat:2015mca} and MMHT14~\cite{Harland-Lang:2014zoa}. These uncertainties are at the level of few to ten percent across the entire mass range except for the heaviest resonances. In particular, when $m_X \gtrsim 3$~TeV, the relative error from NNPDF3.1 quickly becomes $\mathcal{O}(1)$, while the other two collaborations do not exhibit this behavior and their relative errors are under control in the entire mass range of Figs.~\ref{fig:dijet}~and~\ref{fig:dijetNC}. We numerically compare  MMHT14 and NNPDF3.1 sets breaking down the luminosities by the flavor content. The difference between sets is particularly prominent for some flavor combinations, for example, $s \bar c$ fusion, and less relevant for others. Hence, this feature of the NNPDF3.1 set questions the robustness of the limits shown in Figs.~\ref{fig:dijet}~and~\ref{fig:dijetNC} for $m_X \gtrsim 3$~TeV which will be resolved by the future updates.

In the analysis above, we always assume $\mathcal{B}(X\to jj)=1$ as expected in the realistic case for non-leptonic decays given their present sensitivity. This also provides conservative bounds, as any other final state (e.g. leptons) would be easier to detect. The most important caveat concerns the total width of the resonance.  The typical detector resolution of the dijet invariant mass is at the level of $10\%$~\cite{Sirunyan:2019vxa,Sirunyan:2019pnb,Aaboud:2018fzt,Sirunyan:2018xlo,Aad:2019hjw,Sirunyan:2019vgj}. Therefore, a narrow resonance is experimentally defined by $\Gamma_X / m_X \lesssim 10\%$. This criterion has to be satisfied to apply the bounds from Fig.~\ref{fig:dijet}. 

When the resonance $X$ is a bit broader, the dijet limits weaken but do not disappear completely. Fig.~10 of Ref.~\cite{Sirunyan:2019vgj} shows how the cross section times acceptance drops when increasing the total decay width. For example, assuming $m_X=4$~TeV, and increasing the width from $1\%$ to $10\%$ and $40\%$, the limit on {the cross section} 
relaxes by a factor $\sim 3$ and $\sim 15$, respectively. The effect is less pronounced for lighter resonances, for example, when $m_X=2.2$~TeV, these factors are $\sim 1.6$ and $\sim 8$, respectively. To conclude, the limits on the couplings of a broad resonance will be relaxed by at most $\mathcal{O}(1)$ factor from those in Fig.~\ref{fig:dijet}, before the control over the calculation is lost and the strongly-coupled regime is entered.

\section{General EFT analysis of $b \to c u d_i$}
\label{sec:EFT}


In this section we study NP effects in non-leptonic $B$ meson decays {adopting} a bottom-up approach to keep the discussion as general as possible. To this purpose, we utilize the methods of effective field theory (EFT). After presenting the data, we perform the fit in the weak effective Hamiltonian to identity the preferred parameter space. The results are then interpreted in the context of the EFT above the electroweak scale (SMEFT). 

\subsubsection*{Measurements} 

The experimental values for the branching fractions for $b\to c \bar{u} d_i$ decays are obtained by fitting all available data and are compared with the most up-to-date SM predictions based on QCDF~\cite{Bordone:2020gao}. The main difference between the results in Ref.~\cite{Bordone:2020gao} with respect to previous analyses (see e.g. Ref.~\cite{Huber:2016xod}) is the use of updated inputs for CKM elements, decay constants and form factors for $\bar{B}_q\to D^{(*)+}_q$ transitions \cite{Bordone:2019guc,Bordone:2019vic,Zyla:2020zbs}, causing shifts in the central values (the largest one for $\bar{B}_s\to D^{*+}_s \pi$ decay) and generally a reduction of the uncertainties on the branching ratios. From the experimental point of view, the non-leptonic $\bar{B}_s\to D_s^{(*)+} \pi$ and $\bar{B}\to D^{(*)+} K$ decays are often measured as part of ratios with other decay channels in order to reduce experimental errors.

To be conservative we choose to employ the experimental fit in the third column of Table~II of Ref.~\cite{Bordone:2020gao} (without QCDF inputs but with the LHCb measurements of $f_s / f_d$ from semileptonic decays).
Let us define the ratio of the measured branching ratio to the respective SM prediction as
\be
	R(X \to Y Z) \equiv \Br(X \to Y Z) /  \Br(X \to Y Z)_{\SM}~.
	\label{eq:RSMratio}
\ee
Combining the measurements and SM predictions, including correlations in both experimental and theoretical uncertainties,\footnote{We thank Martin Jung for providing the associated correlation matrix.} we obtain the following result:
\be
	\begin{array}{r l}
		R(\bar{B}_s^0 \to D_s^+ \pi^-) &= 0.704 \pm 0.074 \\
		R(\bar{B}^0 \to D^+ K^-) &= 0.687 \pm 0.059\\
		R(\bar{B}_s^0 \to D_s^{*+} \pi^-) &= 0.49 \pm 0.24\\
		R(\bar{B}^0 \to D^{*+} K^-) &= 0.66 \pm 0.13 
	\end{array}~, \qquad
	\rho = \left( \begin{array}{c c c c }
		1 & 0.36 & 0.16 & 0.092 \\
		0.36 & 1 & 0.072 & 0.16 \\
		0.16 & 0.072 & 1 & 0.40 \\
		0.092 & 0.16 & 0.40 & 1
	\end{array}\right)~,
	\label{eq:measurements}
\ee
where $\rho$ is the correlation matrix and, by definition all $R=1$ in the SM.
The observed branching ratios are consistently smaller than the QCDF predictions~\cite{Bordone:2020gao}.

\vspace{10pt}
 
\begin{table}[t]
\centering\small
\begin{tabular}{|l|l|}\hline
$\mathcal{Q}_{V_{LL}}^{ijkl}=(\bar u_L^i \gamma_\mu d_L^j)(\bar d_L^k \gamma^\mu u_L^l) $ &
$\mathcal{Q}_{V_{LL}}^{\prime ijkl}=(\bar u_L^i \gamma_\mu T^A d_L^j)(\bar d_L^k \gamma^\mu T^A u_L^l) $\\[0.2em]
$\mathcal{Q}_{V_{RR}}^{ijkl}=(\bar u_R^i \gamma_\mu d_R^j)(\bar d_R^k \gamma^\mu u_R^l) $ &
$\mathcal{Q}_{V_{RR}}^{\prime ijkl}=(\bar u _R^i \gamma_\mu T^A d_R^j)(\bar d_R^k \gamma^\mu T^A u_R^l)$\\[0.2em]
$\mathcal{Q}_{V_{LR}}^{ijkl}=(\bar u_L^i\gamma_\mu d_L^j)(\bar d_R^k \gamma^\mu u_R^l) $ &
$\mathcal{Q}_{V_{LR}}^{\prime ijkl}=(\bar u_L^i\gamma_\mu T^A  d_L^j)(\bar d_R^k \gamma^\mu  T^A u_R^l) $\\[0.2em]
$\mathcal{Q}_{S_{RL}}^{ijkl}=(\bar u_L^i  d_R^j)(\bar d_R^k u_L^l)$ &
$\mathcal{Q}_{S_{RL}}^{\prime ijkl}=(\bar u_L^i T^A  d_R^j)(\bar d_R^k T^A  u_L^l)$ \\[0.2em]
$\mathcal{Q}_{S_{LR}}^{ijkl}=(\bar u_R^i   d_L^j)(\bar d_L^k  u_R^l)$ &
$\mathcal{Q}_{S_{LR}}^{\prime ijkl}=(\bar u_R^i T^A  d_L^j)(\bar d_L^k  T^A u_R^l)$ \\[0.2em]
$\mathcal{Q}_{S_{RR}}^{ijkl}=(\bar u_L^i   d_R^j)(\bar d_L^k  u_R^l)$ &
$\mathcal{Q}_{S_{RR}}^{\prime ijkl}=(\bar u_L^i T^A  d_R^j)(\bar d_L^k  T^A u_R^l) $ \\[0.2em]
$\mathcal{Q}_{T_{RR}}^{ijkl}=(\bar u_L^i \sigma_{\mu\nu}  d_R^j)(\bar d_L^k  \sigma^{\mu\nu} u_R^l)  $ &
$\mathcal{Q}_{T_{RR}}^{\prime ijkl}=(\bar u_L^i \sigma_{\mu\nu} T^A  d_R^j)(\bar d_L^k  \sigma^{\mu\nu} T^A u_R^l) $ \\[0.2em]
\hline
\end{tabular}
\caption{Low-energy operators relevant for $b \to c \bar u d_i$ transitions.}
\label{tab:Qops}
\end{table}

\subsubsection*{Low-energy effective field theory}

The most general theoretical framework for short-distance NP effects in $b\to c \bar{u} d_i$ ($i=1,2$) transitions is the low-energy effective field theory (LEFT)~\cite{Jenkins:2017jig}. Here we perform a NP analysis including the full set of relevant operators $\mathcal{O}_i$ in the basis of Ref.~\cite{Jenkins:2017jig} (for the list see \eq{eq:LEFT_basis} of Appendix~\ref{app:first}).
These operators, however, are not in a convenient form to evaluate the hadronic matrix elements  we are interested in. Therefore, we Fierz them to the $\mathcal{Q}_i^{(\prime)}$ listed in Table~\ref{tab:Qops}, where
\be
	\mathcal{L}_\text{NP} = \sum_{i=1}^{7} (a_i\,\mathcal{Q}_i + a^\prime_i\,\mathcal{Q}^\prime_i ) + {\rm h.c.}~.
\ee
The corresponding matching relations between the two bases are reported in \eq{eq:matching_LEFT_Q}.
The operators $\mathcal{Q}^{cbiu}_{V_{LL}}$ and $\mathcal{Q}^{\prime cbiu}_{V_{LL}}$ correspond to the SM color-allowed and color-suppressed operators $Q_2$ and $Q_1$ of the CMM basis \cite{Chetyrkin:1997gb}, respectively. In these conventions, the SM Wilson coefficients are
\be
    (a_{V_{LL}}^{cbiu})^{\rm SM} = - C_2 \frac{4 G_F V_{cb} V_{ui}^*}{\sqrt{2}}~, \quad
    (a_{V_{LL}}^{\prime cbiu})^{\rm SM} = - C_1 \frac{4 G_F V_{cb} V_{ui}^*}{\sqrt{2}}~,
\ee
where $C_2 =+1.010$ and $C_1 = -0.291$ \cite{Gorbahn:2004my}.
The hadronic matrix elements for the NP operators are evaluated at leading order in $\alpha_s$ and leading power in $1/m_b$. As in the SM, the NP operators can be grouped in color-allowed and color-suppressed ones. As a consequence of color algebra, we have
\begin{equation}
\langle  D_q^{+(*)}P^-| \mathcal{Q}^\prime_i | \bar{B}_q\rangle = 0 + \mathcal{O}(\alpha_s/N_c)~,
\end{equation}
regardless of the chirality structure of $\mathcal{Q}_i^\prime$ operators. Introducing $\alpha_s$ corrections generates contributions from the color-suppressed operators proportional to $\alpha_s/N_c$.  These contributions in the SM are small compared to the leading ones since they are further suppressed by the Wilson coefficient $C_1 \ll C_2$. A recent computation of color-suppressed topologies \cite{Cai:2021mlt} showed that they are even more subleading than what naively expected, compared to color-allowed ones. This strengthens our hypothesis of disregarding color-suppressed $\mathcal{Q}_i^\prime$ operators for this NP analysis. Furthermore, we stress that this choice does not affect the constraining power of dijet resonance searches.

The non-zero matrix elements give rise to the following decay amplitudes for $\bar{B}_q\to D_q^{+(*)} P^-$ (see Appendix~\ref{app:first} for details):
\be\begin{split}\label{eq:17}
\mathcal{A}(\bar{B}_q\to D_q^{+}P^-) = &\,\mathcal{A}(\bar{B}_q\to D_q^{+}P^-)_\text{SM}\times \\
& \left\{1 +\frac{1}{2\sqrt{2}G_F V_{cb}V^*_{ui} C_2}\bigg[\big(-a_{V_{LL}}^{cbiu} + a_{V_{RR}}^{cbiu} + a_{V_{LR}}^{cbiu}-a_{V_{LR}}^{uibc}\big) \right. \\
&\left.+ \frac{m_P^2}{(m_u+m_{d_i})(m_b-m_c)}\big(a_{S_{RL}}^{cbiu} -a_{S_{LR}}^{cbiu} -a_{S_{RR}}^{cbiu} +a_{S_{RR}}^{uibc} \big)\bigg]\right\}\,, \\
\end{split}\ee
\be\begin{split}
\mathcal{A}(\bar{B}_q\to D_q^{*+}P^-) = &\,\mathcal{A}(\bar{B}_q\to D_q^{*+}P^-)_\text{SM} \times \\
& \left\{1 +\frac{1}{2\sqrt{2}G_F V_{cb}V^*_{ui} C_2}\bigg[\big(-a_{V_{LL}}^{cbiu} - a_{V_{RR}}^{cbiu}+ a_{V_{LR}}^{cbiu}+a_{V_{LR}}^{uibc}\big) \right. \\
&+\left. \frac{m_P^2}{(m_u+m_{d_i})(m_b+m_c)}\big(a_{S_{RL}}^{cbiu}+a_{S_{LR}}^{cbiu}-a_{S_{RR}}^{cbiu}-a_{S_{RR}}^{uibc}\big)\bigg]\right\}\,,
\label{eq:18}
\end{split}\ee
where $i=d,s$ corresponds to $P^- = \pi^-,K^-$, respectively.

\vspace{10pt}

\subsubsection*{Fit to the data} 

These decay amplitudes are used to calculate the ratio of the branching fractions to the respective SM prediction in Eq.~\eqref{eq:RSMratio} as
\be
	R(X \to Y Z) = \frac{|\mathcal{A}(X \to Y Z)|^2}{|\mathcal{A}(X \to Y Z)_{\rm SM}|^2}~,
\ee
from which we perform a fit using Eq.~\eqref{eq:measurements} and  \eqs{eq:17}{eq:18}.
Only half of the NP coefficients contributing to the amplitudes in \eq{eq:17} and \eq{eq:18} can simultaneously explain the observed suppression in both final states with a $D_q^+$ or a $D_q^{+*}$ meson. These are $a^{cbiu}_{V_{LL}}$, $a^{cbiu}_{V_{LR}}$, $a^{cbiu}_{S_{RR}}$ and $a^{cbiu}_{S_{RL}}$. The other half can not fit the data well. Regarding the flavor structure, new interactions with both strange and down quarks are needed. 

\begin{figure}[t]
\centering
\includegraphics[height=6.5cm]{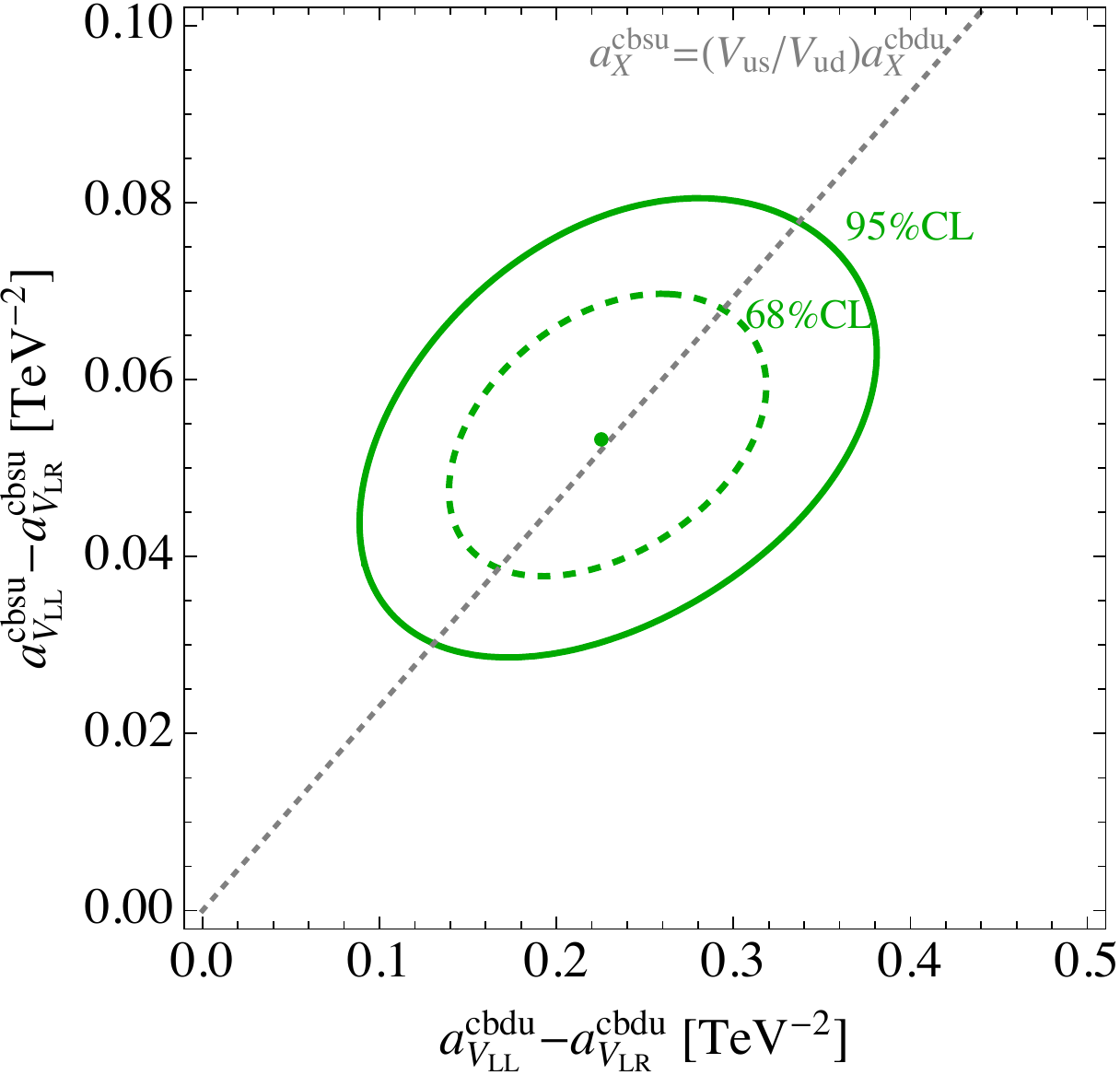} \; \;
\includegraphics[height=6.5cm]{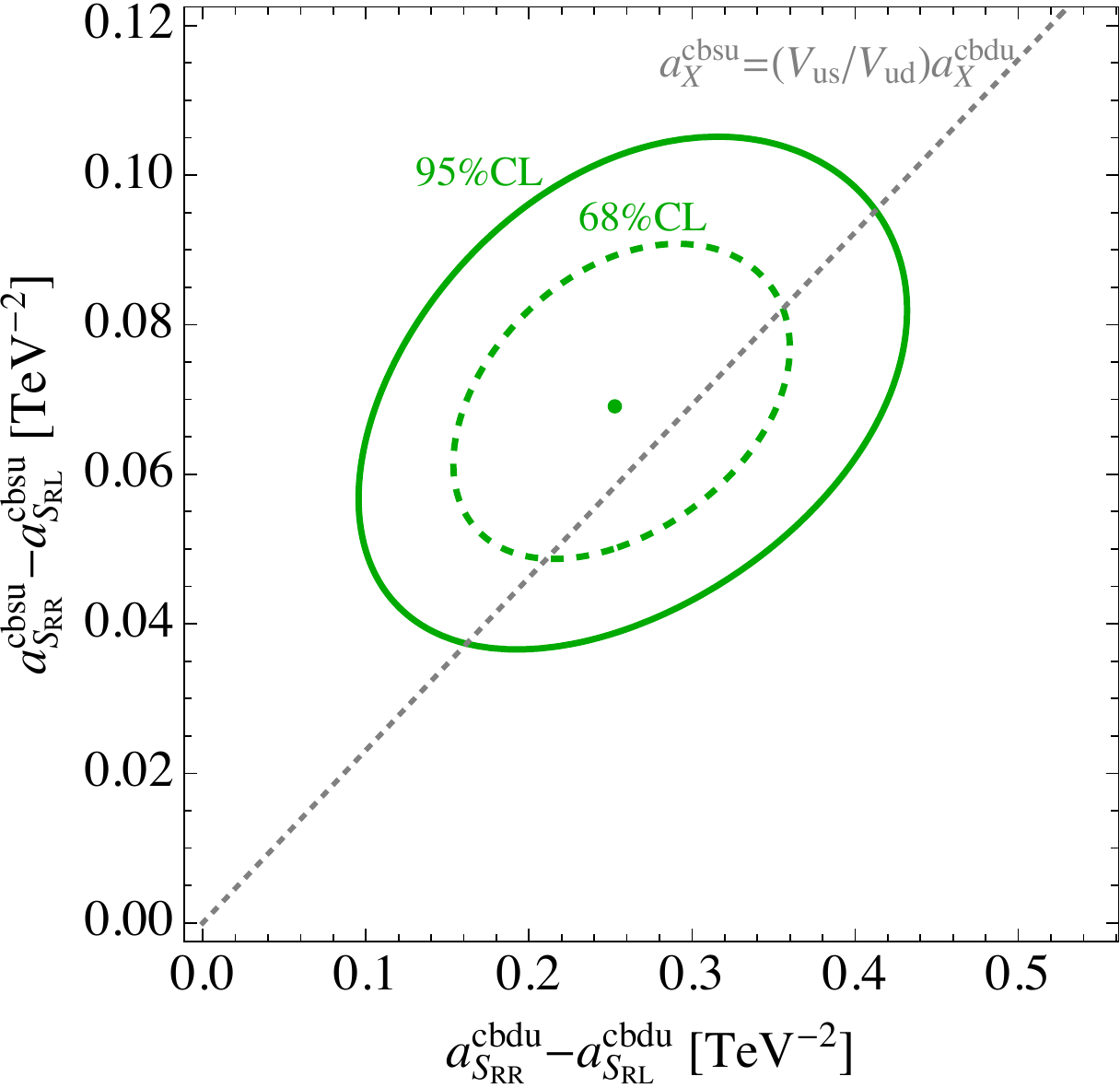}
\caption{\label{fig:fitas} Low-energy EFT fit to $\bar{B}_q\to D_q^{+(*)} P^-$ decays. Dashed and solid lines show $68\%$ and $95\%$ CL regions for vector operators ({\bf left panel}) and scalar operators ({\bf right panel}). The gray dotted line is consistent with the relative size following the CKM ratio $V_{u s}/V_{u d}$.
}
\label{fig:1}
\end{figure}

The results of the fits are shown in \fig{fig:1}. The left (right) panel is for new vector (scalar) interactions. The horizontal and the vertical axes are for the couplings to down and to strange quarks, respectively. The dashed and solid green lines describe the boundaries of the $68\%$ and $95\%$ CL regions. The discrepancy between SM predictions and measurements manifests as a shift of the preferred region from the origin.
The best-fit point in the left (right) plot improves the fit to data with respect to the SM by $\chi^2_{\rm SM} - \chi^2_{\rm best-fit} \approx 36 \, (35)$ and corresponds to
\be\begin{split} \label{eq:fit}
\text{vector:}~\Big\{&	a_{V_{LL}}^{cbdu} - a_{V_{LR}}^{cbdu} \approx 0.23 V_{ud} \TeV^{-2} ~, \quad
		a_{V_{LL}}^{cbsu} - a_{V_{LR}}^{cbsu} \approx 0.24 V_{us}  \TeV^{-2} ~, \\
\text{scalar:}~\Big\{&	a_{S_{RR}}^{cbdu} - a_{S_{RL}}^{cbdu} \approx 0.26 V_{ud}  \TeV^{-2} ~, \quad
		a_{S_{RR}}^{cbsu} - a_{S_{RL}}^{cbsu} \approx 0.31 V_{us}  \TeV^{-2} ~.
\end{split}\ee
The preferred size of the effective operators suggests ultraviolet completion not far above the TeV scale. 
Furthermore, we observe that in both cases the fits are compatible with a CKM-like flavor structure, with the operators involving the strange quark being Cabibbo-suppressed with respect to those with the down quark, as shown with the gray dotted lines in Fig.~\ref{fig:1}. This is a desirable trait from the flavor model building perspective.

\vspace{10pt}

\subsubsection*{Standard model effective field theory}

The EFT coefficients in \fig{fig:1} are reported at scale $\mu_R = m_b$. To establish connections with possible UV completions, these results have to be appropriately extrapolated to high energies. The low-energy EFT coefficients are evolved up to the EW scale and then matched at tree-level to the SMEFT. These are finally evolved to the UV scale (see App.~\ref{sec:SMEFT_ops} for details). In the SMEFT, the theory is supplement with a series of gauge-invariant irrelevant operators of increasing canonical dimension.
Among all possible dimension-six SMEFT coefficients, we focus on the dimension-six four-fermion operators that either contribute directly at tree-level to $b\to c \bar{u} d_i$ or strongly mix with such operators. In Table~\ref{tab:Qops_SMEFT}, we list all these operators. Other tree-level effects in the SMEFT, such as $W$-vertex corrections, are better constrained elsewhere, and can not give sizable effect to $\bar B_q \to D_q^{(*)+} \{\pi, K\}$ decays.

\begin{table}[t]
\centering\small
\begin{tabular}{|l|l|}\hline
$[\mathcal{O}_{qq}^{(1)}]_{ijkl}= (\bar q_L^i \gamma_\mu q_L^j)(\bar q_L^k \gamma_\mu q_L^l)$ & 
$[\mathcal{O}_{qq}^{(3)}]_{ijkl}= (\bar q_L^i \sigma^a \gamma_\mu q_L^j)(\bar q_L^k \sigma^a \gamma_\mu q_L^l)$ \\[0.2em]
$[\mathcal{O}_{ud}^{(1)}]_{ijkl}= (\bar u_R^i \gamma_\mu u_R^j)(\bar d_R^k \gamma_\mu d_R^l)$ &  
$[\mathcal{O}_{ud}^{(8)}]_{ijkl}= (\bar u_R^i T^A \gamma_\mu u_R^j)(\bar d_R^k T^A \gamma_\mu d_R^l)$\\[0.2em]
$[\mathcal{O}_{qd}^{(1)}]_{ijkl}=  (\bar q_L^i \gamma_\mu q_L^j)(\bar d_R^k \gamma_\mu d_R^l)$ &  
$[\mathcal{O}_{qd}^{(8)}]_{ijkl}=  (\bar q_L^i T^A \gamma_\mu q_L^j)(\bar d_R^k T^A \gamma_\mu d_R^l)$ \\[0.2em]
$[\mathcal{O}_{qu}^{(1)}]_{ijkl}=  (\bar q_L^i \gamma_\mu q_L^j)(\bar u_R^k \gamma_\mu u_R^l)$ &  
$[\mathcal{O}_{qu}^{(8)}]_{ijkl}=  (\bar q_L^i T^A \gamma_\mu q_L^j)(\bar u_R^k T^A \gamma_\mu u_R^l)$ \\[0.2em]
$[\mathcal{O}_{quqd}^{(1)}]_{ijkl}=  (\bar q_L^i u_R^j) (i\sigma^2) (\bar q_L^k d_R^l)$ &  
$[\mathcal{O}_{quqd}^{(8)}]_{ijkl}=  (\bar q_L^i T^A u_R^j) (i\sigma^2) (\bar q_L^k T^A  d_R^l)$ \\[0.2em]
\hline
\end{tabular}
\caption{SMEFT operators relevant for $b \to c \bar u d_i$ transitions.}
\label{tab:Qops_SMEFT}
\end{table}

\section{Simplified models}
\label{sec:mediators_dijet}
The SMEFT operators identified in the previous section can be generated already at tree-level by integrating out a new bosonic field $X$ coupled to quark currents. Here we list the complete set of new scalar and vector mediators which generate the relevant operators shown in Fig.~\ref{fig:fitas} at tree-level with renormalisable interactions~\cite{deBlas:2017xtg}, without also necessarily inducing dangerous $\Delta F = 2$ transitions at tree-level
\be\begin{split}
\!\!\!\!\text{spin-0:}  &\quad \left\{ \begin{array}{lll}
        \Phi_1 = ({\bf 1}, {\bf 2}, 1/2) , ~~&
        \Phi_8 = ({\bf 8}, {\bf 2}, 1/2), ~~& \\
		\Phi_3 = ({\bf \bar{3}}, {\bf 1}, 1/3), ~~& 
		\Psi_3 = ({\bf \bar{3}}, {\bf 3}, 1/3), ~~&
		\Phi_6 = ({\bf 6}, {\bf 1}, 1/3), 
	\end{array} \right. \\
\!\!\!\!\text{spin-1:} &\quad \left\{ 
		\mathcal{Q}_3 = ({\bf 3}, {\bf 2}, 1/6), \quad 
		\mathcal{Q}_6 = ({\bf \bar{6}}, {\bf 2}, 1/6)
	\right. ~.
\end{split}
\label{eq:mediators}
\ee
Here, the SM gauge representations are reported in the format $(SU(3)_c, SU(2)_L, U(1)_Y)$. Among other mediators that generate at tree-level the effective operators listed in Eq.~\eqref{eq:SMEFT_basis}, colored vectors $({\bf 3}, {\bf 2}, -5/6)$ and $({\bf \bar{6}}, {\bf 2}, -5/6)$ are not viable since the coefficients $a_{S_{LR}}^{cb i u}$ do not fit the anomaly. On the other hand, the vector triplet $W^\prime =  ({\bf 1}, {\bf 3}, 0)$, vectors $({\bf 8}, {\bf 1}, 0)$ and $({\bf 8}, {\bf 3}, 0)$, and the scalar $({\bf 6}, {\bf 3}, 1/3)$ mediate a neutral meson mixing at tree-level even with the minimal set of couplings required to fit the anomaly. Hence, we do not consider them further given the stringent constraints on $\Delta F = 2$ transitions.  We refer to Ref. \cite{Iguro:2020ndk} for a more detailed discussion of the $W^\prime$ case.

The main goal of this section is to show how the high-$p_T$ searches at the LHC, specifically those from dijet signatures, can test the solutions of the anomaly for all viable mediators. We separate the discussion into two subsections based on the pair production dijet resonance searches from \sec{sec:highpTpair}. In particular, in \sec{sec:coloredM} we focus on colored resonances which receive important constraints from the pair production at the LHC, while the colorless doublet $\Phi_1$ is studied in isolation in \sec{sec:Higgs}. The colorless mediator can in principle be much lighter since the relevant bound comes only from the LEP-II collider.

The single dijet resonance searches derived in Section~\ref{sec:highpTsingle} can be used in both cases. Non-leptonic meson decays depend on the product of two couplings when the resonance is integrated out at tree level. In particular, the product of the couplings entering those decays satisfies 
\begin{equation}
| x_{q^i q^j} \,x^*_{q^k q^l} | = |x_{q^i q^j}| \times |x_{q^k q^l}|~,\label{eq:inequality}
\end{equation}
where both terms on the right-hand side are simultaneously constrained from non-observation of $\sigma(p p \to X \to j j )$ at high-$p_T$. Using this inequality, we can limit NP contributions in $\bar B_q\to D_q^{(*)+}  \,\{\pi, K\}$ decays.\footnote{We only consider scenarios in which the anomaly is attributed to a single new mediator. One could explore a possibility of several mediators, each passing the dijet bounds, while adding up in $\bar B_q\to D_q^{(*)+}  \,\{\pi, K\}$.}

%

\subsection{Colored mediators}
\label{sec:coloredM}

\begin{figure}[t]
\centering
\includegraphics[height=5.3cm]{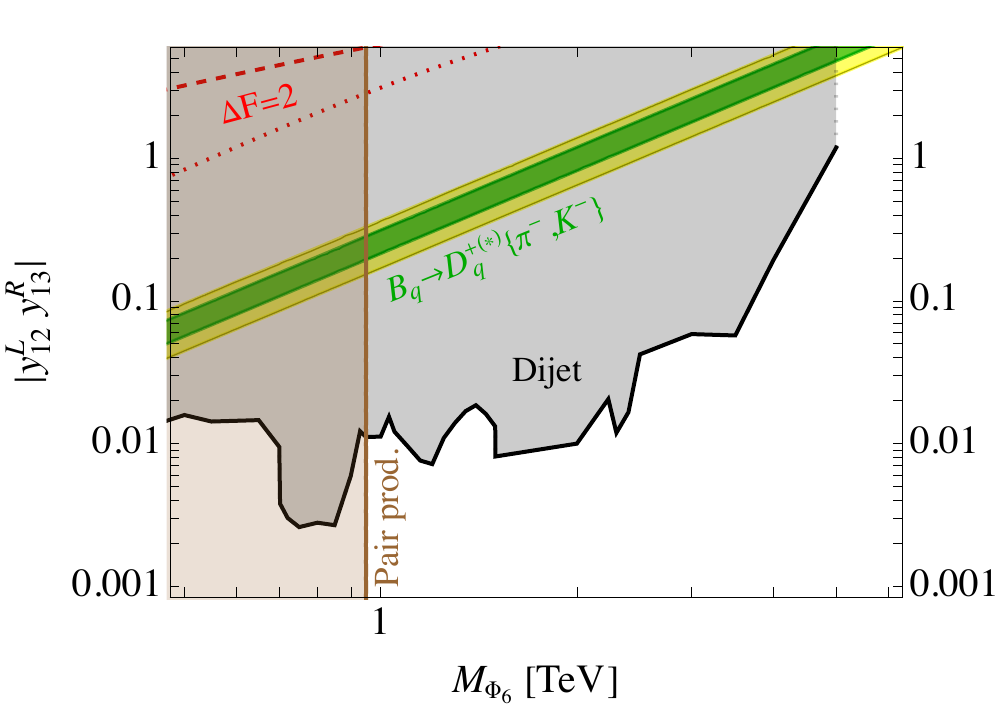} 
\includegraphics[height=5.3cm]{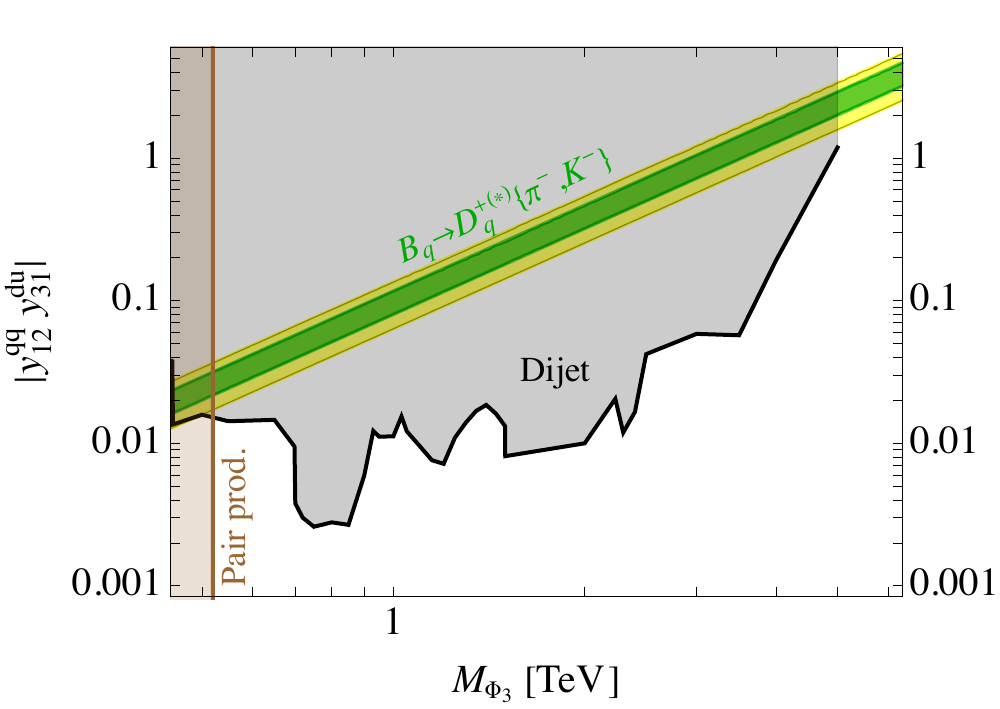} \\
\includegraphics[height=5.3cm]{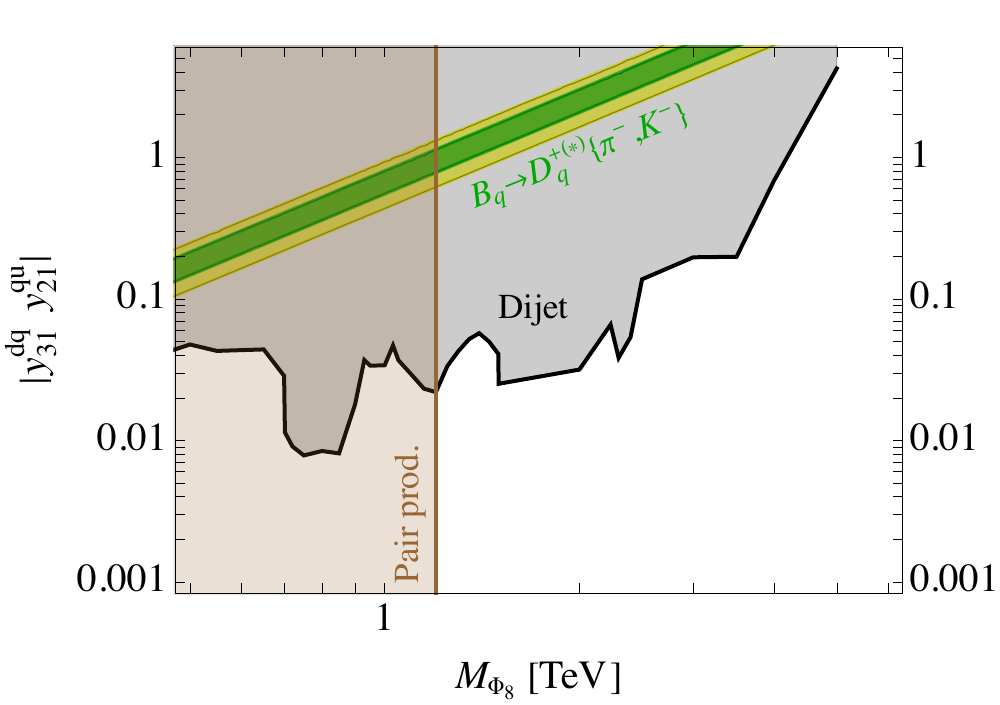} 
\includegraphics[height=5.3cm]{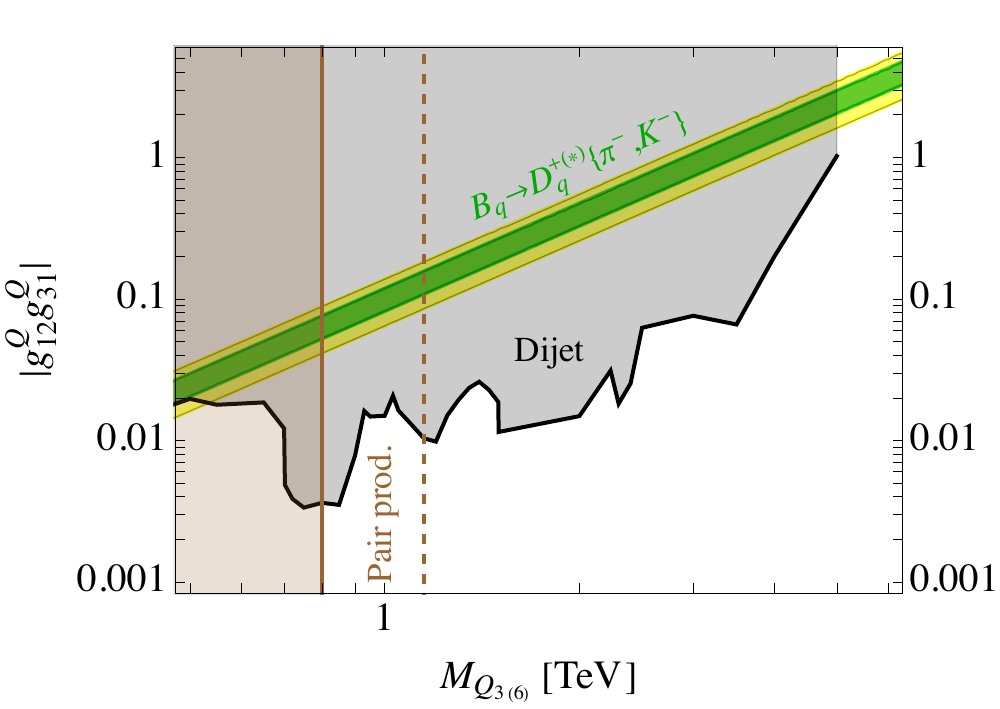} 
\caption{\label{fig:colored_vs_dijet} High-$p_T$ constraints from the single dijet resonance production (gray) and the QCD-induced pair production of dijet resonances (brown) compared with the best-fit region from non-leptonic $B$ decays. The constraints are imposed on the product of the two relevant couplings as a function of the mass for colored mediators listed in Eq.~\eqref{eq:mediators}. Note that the constraints from \sec{sec:highpTsingle} are strictly applicable for $\Gamma_X / m_X \lesssim 10\%$ which is not necessarily the case in the upper parts of the plot, depending on the relative sizes of the couplings. The anomaly in $\bar{B}_q\to D_q^{+(*)} P^-$ selects the best-fit region at 68\% CL (green) and 95\% CL (yellow). Shown with the red dashed lines in the top-left plot ($\Phi_6$) are the limits from the meson mixing for two representative choices of $y_{1 3}^R$ coupling. For more details see \sec{sec:coloredM}.}
\end{figure}

As discussed in \sec{sec:highpTpair}, the QCD-induced pair production at the LHC sets robust lower limits on the masses of the colored mediators in the range 0.5~TeV to 1.15~TeV, depending on the representation. Note that, complementary to the pair production, the single mechanism is effective for heavy resonances. The combination of single and pair production dismisses all these mediators as the explanation of the anomaly, see Fig.~\ref{fig:colored_vs_dijet}. 

In the following we show the interplay between the dijet bounds and the fit to the anomaly for each mediator, leaving the details on the models and their EFT matching to App.~\ref{sec:mediators}. 

\subsubsection*{Color-sextet diquark $\Phi_6$}

The SM is extended with the singlet sextet diquark scalar $\Phi_6 = ({\bf 6}, {\bf 1}, 1/3)$.
The relevant interaction Lagragian is:
\be
	\LL_{\Phi_6} \supset y^L_{ij}  {\bf \Phi}_6^{\alpha\beta \dagger} \bar{q^c}^{(\alpha|}_{L i} (i \sigma_2) q^{|\beta)}_{Lj} + y^R_{ij} {\bf \Phi}_6^{\alpha\beta \dagger} \bar{u^c}^{(\alpha|}_{R i} d^{|\beta)}_{Rj} + {\rm h.c.}~,
\ee
where $\psi_i^{(\alpha|} \psi_j^{|\beta)} = \frac{1}{2} (\psi_i^{\alpha} \psi_j^{\beta} + \psi_i^{\beta} \psi_j^{\alpha})$, and $y^L$ is an antisymmetric matrix. The components of the sextet representations are given as ${\bf \Phi}_6^{\alpha\beta} \equiv S^{i}_{\alpha\beta} \Phi_6^i$, where $i = 1, \ldots, 6$ and the symmetric color matrices $S^i_{\alpha\beta}$ are given in Eq.~\eqref{eq:SymmColMatr}. The anomaly can be addressed by switching on only two couplings:
\be
	y^{L} = \left( \begin{array}{ccc}
	0 	&	y^L_{12} & 0 \\
	- y^L_{12} & 0 & 0 \\
	0 & 0 & 0
	\end{array}\right)~, \qquad
	y^{R} = \left( \begin{array}{ccc}
	0 & 0 & y^R_{13} \\
	0 & 0 & 0 \\
	0 & 0 & 0
	\end{array}\right)~.
	\label{eq:S6Coupl}
\ee
It is worth noticing that the structure of the left-handed couplings $y^L$ of Eq.~\eqref{eq:S6Coupl} is compatible with the approximate $U(2)_q$ symmetry of the SM Lagrangian, where the first two families transform as a doublet while the third as a singlet \cite{Barbieri:2011ci}. Indeed, since the antisymmetric combination of the $q_L^{i=1,2}$ doublets transforms as a singlet, the $U(2)_q$ symmetry would predict $y^L_{12} \sim \mathcal{O}(1)$ while $y^L_{13}, y^L_{23} \ll 1$.
Regarding the right-handed couplings, the $U(2)_u \times U(2)_d$ symmetry would predict $y^R_{33} \sim 1$ while all other terms should be suppressed.
By introducing a spurion $V_u = (a_u, 0)$, with $a_u \ll 1$, transforming as a doublet of $U(2)_u$ it is possible to generate a small value of $y^R_{13} \sim a_u \ll 1$. This spurion is not required by the minimal breaking of the symmetry necessary to generate the SM Yukawas. We thus conclude that this setup could be compatible with a non-minimally broken $U(2)^5$ flavor symmetry if $y^L_{12} \sim 1$, $y^R_{33} \sim 1$ and $y^R_{13} \ll 1$.

In the following, we considering the minimal set of couplings introduced in Eq.~\eqref{eq:S6Coupl}. The non-vanishing $a_X$ coefficients for $\bar{B} \to D_q^{(*)} P^-$ decays are
\be
	a_{S_{RR}}^{cbdu} \approx \frac{2}{3} \kappa_{\rm RGE}^{S} V_{c s} \frac{  y^{L *}_{12} y^{R}_{13}}{ M_{\Phi_6}^2} \approx \frac{0.26 V_{cs}}{\TeV^2}~, \qquad
	a_{S_{RR}}^{cbsu} \approx - \frac{2}{3} \kappa_{\rm RGE}^{S} V_{c d} \frac{  y^{L *}_{12} y^{R}_{13}}{ M_{\Phi_6}^2} \approx \frac{-0.31 V_{cd}}{\TeV^2} ~,\label{eq:6fit}
\ee
where $\kappa_{\rm RGE}^{S} \approx 1.65 \, (1.85) $ for $M_{\Phi_6} = 1 \, (5) \TeV$. The $1\sigma$ and $2\sigma$ regions from the anomaly fit in the plane of the product of the two couplings and the mediator mass is shown as a green and yellow band in the top-left panel of Fig.~\ref{fig:colored_vs_dijet}, respectively.
This state contributes to precisely measured flavor-violation processes at one-loop level. We study the relevant constraints in \app{sec:SextetFlavor}, which place an upper limit on $y^L_{12}$ as function of the mass. The stronger bound comes from $D^0$ and $K^0$ mixing, and the limit is shown in Fig.~\ref{fig:colored_vs_dijet} (top-left) as dashed and dotted red lines, for two different assumptions on the right-handed coupling $y^R_{13}$. As shown in the plot, these complementary constraints from low-energy measurements are not able to probe the interesting parameter space.

The dijet resonance limits become less effective when the resonance is broader,
\be
    \frac{\Gamma_{\Phi_6} }{ M_{\Phi_6} } = \frac{8 |y^L_{12}|^2 + |y^R_{13}|^2}{16 \pi} \gtrsim 0.1 ~,
\ee
but the perturbativity of the model comes into question. 
Fixing the best-fit value for the product of the two couplings, the condition above is always violated for $M_{\Phi_6} \gtrsim 2$~TeV. The strongest dijet constraints on the two couplings $y^L_{12}$ and $y^R_{13}$ entering Eq.~\eqref{eq:6fit}, arise from the processes $u s \to \Phi_6$ and $u b \to \Phi_6$, respectively. The product of the two couplings, that contributes directly to the anomaly, is bounded since $|y^L_{12} y^R_{13}| < |y^L_{12}|^{\rm max} |y^R_{13}|^{\rm max}$.
Compared to the generic Lagrangian in Eq.~\eqref{eq:X_lagrangians} the couplings are given by $x_{ij} = 2 y^L_{ij}$ for the left-handed quarks and $x_{ij} = y^R_{ij}$ for the right-handed ones.
As shown in Fig.~\ref{fig:colored_vs_dijet} (top-left), the region preferred by the anomaly is excluded by the dijet searches for all masses where the theory is perturbative.

\subsection*{Color-triplet diquark $\Phi_3$}

The scalar triplet $\Phi_3 = ({\bf \bar 3}, {\bf 1}, 1/3)$ couples to the SM quarks as
\be
	\LL_{\Phi_3} \supset y^{qq}_{ij} \epsilon_{\alpha\beta\gamma} \Phi_3^\alpha \bar{q}^\beta_{Li} (i \sigma_2) q^{c\, \gamma}_{Lj} + y^{du}_{ij} \epsilon_{\alpha\beta\gamma} \Phi_3^\alpha \bar{d}^\beta_{Ri} u^{c\, \gamma}_{Rj} + {\rm h.c.}~,
\ee
where $y^{qq}_{ij}$ is a symmetric matrix. Baryon number conservation must be imposed to suppress the couplings to quark and leptons, that would otherwise mediate proton decay. The coupling structure that allows fitting the anomaly with least suppression demands three non-vanishing couplings $y^{du *}_{31}$, $y^{qq}_{12}$, and $y^{qq}_{22}$, such that
\be
	a_{S_{RR}}^{c b d u} = - 2.6 \frac{ y^{qq}_{12} y^{du *}_{31} }{ M_{\Phi_3}^2} \approx \frac{0.26 V_{ud}}{\TeV^2}, \quad
	a_{S_{RR}}^{c b s u} = \frac{ (- 2.6 y^{qq}_{22} + 0.60 y^{qq}_{12} ) y^{du *}_{31}}{ M_{\Phi_3}^2} \approx \frac{0.31 V_{us}}{\TeV^2}.
\ee
A good fit requires the relation $y^{qq}_{22} \approx 0.50 y^{qq}_{12}$.

The partonic processes that give the strongest constraints on the couplings relevant to this model are the same as in the scalar sextet case, as well as the the relations between the $y^{qq/du}_{ij}$ and $x_{ij}$ couplings. As shown in Fig.~\ref{fig:colored_vs_dijet} (top-right), the dijet searches firmly exclude the parameter space relevant for the anomaly in all the perturbative range of the model.

Potentially strong limits from loop-induced flavor-violating processes might require a particular coupling structure. We do not discuss them further since the dijet searches are already quite restrictive. The case of the scalar $\Psi_3 = ({\bf \bar 3}, {\bf 3}, 1/3)$ is discussed in the appendix and shares analogous features as the scenario where $\Phi_3$ only couples to LH quarks. This scenario is not so advantageous for the anomaly since it involves a sizable coupling to the top quark, which implies stronger collider constraints. For this reason, we do not consider it separately, referring to App.~\ref{sec:mediators} for more details.

\subsection*{Color-octet scalar $\Phi_8 = (8, 2, 1/2)$}

The scalar octet $\Phi_8$ couples to quarks with the Lagrangian
\be
	\LL_{\Phi_8} \supset y^{qu}_{ij} \Phi_8^{\alpha \dagger} i \sigma_2 \bar{q}^T_{L i} T^\alpha u_{R j} + y^{dq}_{ij} \Phi_8^{\alpha \dagger} \bar{d}_{R i} T^\alpha q_{L j}  + {\rm h.c.}~.
\ee
In order to fit the anomaly with minimal CKM suppression and least possible effect in dijet searches we consider the following non-vanishing couplings: $y^{ dq}_{31}$, $y^{ dq}_{32}$, and $y^{qu}_{21}$. The low-energy coefficients induced by these {couplings} are:
\be
	a_{S_{RR}}^{c b i u} \approx 0.44 V_{c s} \frac{ y^{ dq *}_{3 i} y^{ qu}_{2 1}}{ M_{\Phi_8}^2}~, 
\ee
where $i = 1,2$. We take $y^{ dq *}_{3 2} = V_{us}/V_{ud} ~ y^{ dq *}_{3 1}$ in order to be consistent with the relative effect observed in $K$ and $\pi$ channels.
In the case of the color-octet representation, the limits from QCD pair production are particularly strong, forbidding masses below $\sim 1$~TeV. Furthermore, the relatively small numerical factor in the low-energy coefficients $a_{S_{RR}}^{c b i u}$ requires larger couplings to fit the anomaly compared to the previous models. These facts, combined, exclude a successful explanation of the observed deviation with this setup, as shown in  Fig.~\ref{fig:colored_vs_dijet} (bottom-left).

\subsubsection*{Vectors $\QQ_3$ and $\QQ_6$}

The triplet and sextet vectors $\QQ_3 = ({\bf 3}, {\bf 2}, 1/6)$ and $\QQ_6 = (\bar {\bf 6}, {\bf 2}, 1/6)$ interact with SM quarks as
\be
	\LL_{\QQ} \supset g^{\QQ_3}_{ij}  \QQ_3^{\alpha \mu \dagger} \epsilon_{\alpha\beta\gamma} \bar{d}^{\beta}_{R i} \gamma_\mu (i \sigma_2) q^{c \gamma}_{Lj} + \frac{1}{2} g^{\QQ_6}_{ij}  \QQ_6^{\alpha\beta \mu \dagger} \bar{d}^{(\alpha|}_{R i} \gamma_\mu (i \sigma_2) q^{c |\beta)}_{Lj}  + {\rm h.c.}~.
\ee
The tree-level contribution to the low-energy EFT coefficients relevant for the anomalies is given by
\be
	a_{S_{RL}}^{c b i u} = \frac{4}{3} \kappa_{\rm RGE}^{S} V^*_{u i} V_{cj} \left(\frac{ g^{\QQ_3 *}_{3i} g^{\QQ_3}_{i j} }{ M_{\QQ_3}^2} - \frac{ g^{\QQ_6 *}_{3i} g^{\QQ_6}_{i j} }{ M_{\QQ_6}^2} \right)~,
\ee
where $\kappa_{\rm RGE}^{S} \approx 2.23$ for a scale of 1~TeV. The two states give the same contribution, up to a sign change for one coupling. The combination which has a weaker CKM suppression is obtained with these three couplings only: $g^{\QQ, dq}_{31}$, $g^{\QQ, dq}_{12}$, and $g^{\QQ, dq}_{22}$. In particular,
\be
	a_{S_{RL}}^{c b d u} \approx \frac{3.0 V_{cs} V_{ud}^*}{M_{\QQ}^2} g^{\QQ *}_{31} g^{\QQ}_{12} \approx - \frac{0.26 V_{ud}}{\TeV^2}~, \quad
	a_{S_{RL}}^{c b s u} \approx \frac{3.0 V_{cs} V_{ud}^*}{M_{\QQ}^2} g^{\QQ *}_{31} g^{\QQ}_{22}  \approx - \frac{0.31 V_{us}}{\TeV^2}~.
\ee
To fit the anomaly, we impose the relation $g^{\QQ}_{22} = g^{\QQ}_{12} V_{us}^*/V_{ud}^*$.

In this scenario the leading partonic processes for dijet production are $d c \to \QQ_3^{-1/3 *}$, $d u \to \QQ_3^{-1/3 *}$ (Cabibbo-suppressed), and $d s \to \QQ_3^{2/3 *}$, induced by $g^{\QQ}_{12}$, and  $b u \to \QQ_3^{-1/3 *}$ and $b d \to \QQ_3^{2/3 *}$, induced by $g^{\QQ}_{31}$. The strongest limits are from the $d s$ ($d u$ in the high-mass region) and $b u$ induced ones.
Also, in this case, the dijet constraints exclude a weakly-coupled solution of the anomaly, see Fig.~\ref{fig:colored_vs_dijet} (bottom-right).

\subsection{Colorless scalar doublet model}
\label{sec:Higgs}
	
The scalar doublet $\Phi_1 = ({\bf 1}, {\bf 2}, 1/2)$ is among the possible tree-level mediators capable to fit the anomaly, Eq.~\eqref{eq:mediators}. It is, however, a unique mediator in the list since it is not charged under QCD and therefore not sufficiently constrained by the pair production at the LHC. For generic Yukawa couplings, its neutral component will mediate $\Delta F = 2$ transitions at tree-level. This can be avoided by a suitable alignment in flavor space. While such alignment is theoretically rather unappealing, it is still motivated to consider this option as it opens up a qualitatively different region of parameter space where the mediator mass is comparable to the heaviest particles in the SM. In the case of colored mediators discussed above, $p p \to X X \to (jj) (jj)$ searches imply a mass gap from the SM states, and to fit the anomaly, this means larger couplings, such that $p p \to X \to jj$ becomes important. For this reason, we study the simplified $\Phi_1$ model in details.

Having the same quantum numbers as the SM Higgs boson, the two states will mix in general. This would disrupt the precise flavor alignment required to pass the meson mixing constraints and must be forbidden. For the sake of this simplified analysis, we just assume that $\Phi_1$ is the mass eigenstate corresponding to the doublet which does not take a vacuum expectation value and that no mixing is present at tree-level. Regarding its Yukawa couplings, we consider two different benchmark scenarios, designed ad-hoc to avoid tree-level contributions to meson mixing:

{\bf Benchmark I} --- The couplings of the extra scalar $\Phi_1$ are exclusively to the right-handed down quarks and are diagonal in the down-quark mass basis,
\begin{equation}
	\mathcal{L}_{\Phi_1}^{\rm Yuk} = y^d_{i} \, \Phi_1^{\dagger}\bar d^i_R  q^i_L  \,  + {\rm h.c.},
\end{equation}
where $q^i_{L} = (V^*_{j i} u^j_L, d^i_L)^T$. A possible mechanism behind this alignment is discussed in~\cite{Egana-Ugrinovic:2018znw,Egana-Ugrinovic:2019dqu,Egana-Ugrinovic:2021uew}.
Integrating out the scalar $\Phi_1$, the LEFT operators $L_{ud}^{V1(8),LR}$ are generated at low energies, which contribute to the $a_{S_{RL}}^{ijkl}$ coefficients as
\begin{equation}
a_{S_{RL}}^{cbiu} = \kappa_{\rm RGE}  V_{cb} V^*_{ui} \frac{y^{d *}_3 y^{d}_i}{M^2_{\Phi_1}}~,
\end{equation}
where $\kappa_{\rm RGE} \approx 2.0$ for $M_{\Phi_1} = 200\,\GeV$, derived using DsixTools 2.0 \cite{Fuentes-Martin:2020zaz}.
This structure is compatible with the fit in the right panel of Fig.~\ref{fig:fitas}, where the relation $y^d_1 = y^d_2$ allows to simplify the analysis.

{\bf Benchmark II} --- The couplings of $\Phi_1$ are aligned to the right-handed bottom quark and to the right-handed up quark:
\begin{equation}
\mathcal{L}_{\Phi_1}^{\rm Yuk} = y^d_3 \, \Phi_1^{\dagger} \bar b_R  q^3_L \, + y^u_1 \, \bar{\tilde q}^1_L u_R   \, \tilde{\Phi}_1 + {\rm h.c.},
\end{equation}
where $q^3_{L} = (V^*_{j b} u^j_L, b_L)^T$, ${\tilde q}^1_{L} = (u_L, V_{u j} d^j_L)^T$, and $y^u_1$ and $y^d_3$ are complex numbers.
In this setup the relevant LEFT coefficients generated at low energies are $L_{uddu}^{S1,RR}$, which contribute as
\begin{equation}
a_{S_{RR}}^{cb iu} = \kappa_{\rm RGE} V_{cb} V_{ui} \frac{y^{d*}_3 y^{u}_1}{M^2_{\Phi_1}}~,
\end{equation}
where $\kappa_{\rm RGE} \approx 2.07$ for $M_{\Phi_1} = 200$\,GeV.
Also this benchmark fits very well the excess in the hadronic $B$ decays.

\begin{figure}[t]
\centering
\includegraphics[width=6cm]{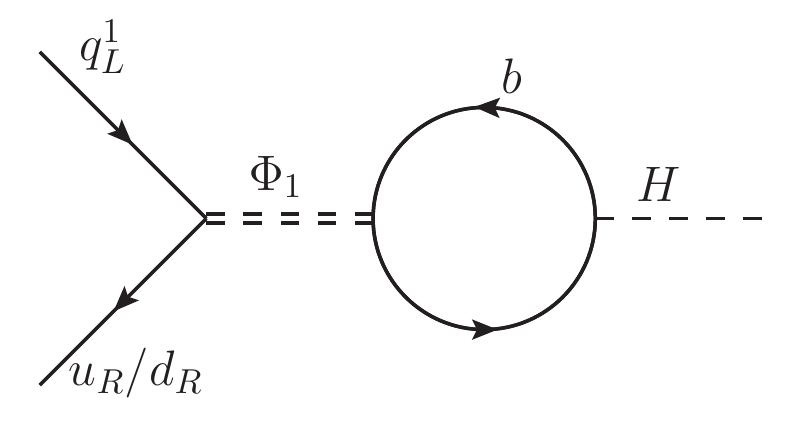}
\caption{\label{fig:Hmix} Radiative corrections to the SM Higgs Yukawa couplings from the desired $\Phi_1$ interactions.}
\end{figure}

Let us now briefly discuss the stability of this alignment under radiative corrections.
The couplings of $\Phi_1$ with fermions induce at one loop mixing with the SM Higgs. With the coupling structures outlined above, the largest contribution in both benchmarks arise from the bottom-quark loop, giving a mixing angle $\sim y^d_3 Y_b/(16\pi^2) \sim \mathcal{O}(10^{-4})$.
In Benchmark I (II) this induces a correction to the down (up) quark Yukawa, as in Fig.~\ref{fig:Hmix}, with an estimated size of
\be
    \delta Y_{d(u)} \sim \frac{y^d_1 (y^u_1) y^d_3 Y_b }{16\pi^2} \frac{m_H^2}{M_{\Phi_1}^2} \sim ~ 0.3 Y_d (0.6 Y_u),
\ee
where we used the best-fit values from the fits to the anomaly.
Due to the small expected size of this effect and the $\Phi_1$ couplings being flavor-diagonal, no fine tuning against radiative corrections is required to keep this alignment and to avoid tree-level FCNC. Nonetheless, the odd flavor alignments still remain to be justified.

\subsubsection*{Collider constraints}

\begin{figure}[t]
\centering
\includegraphics[height=7cm]{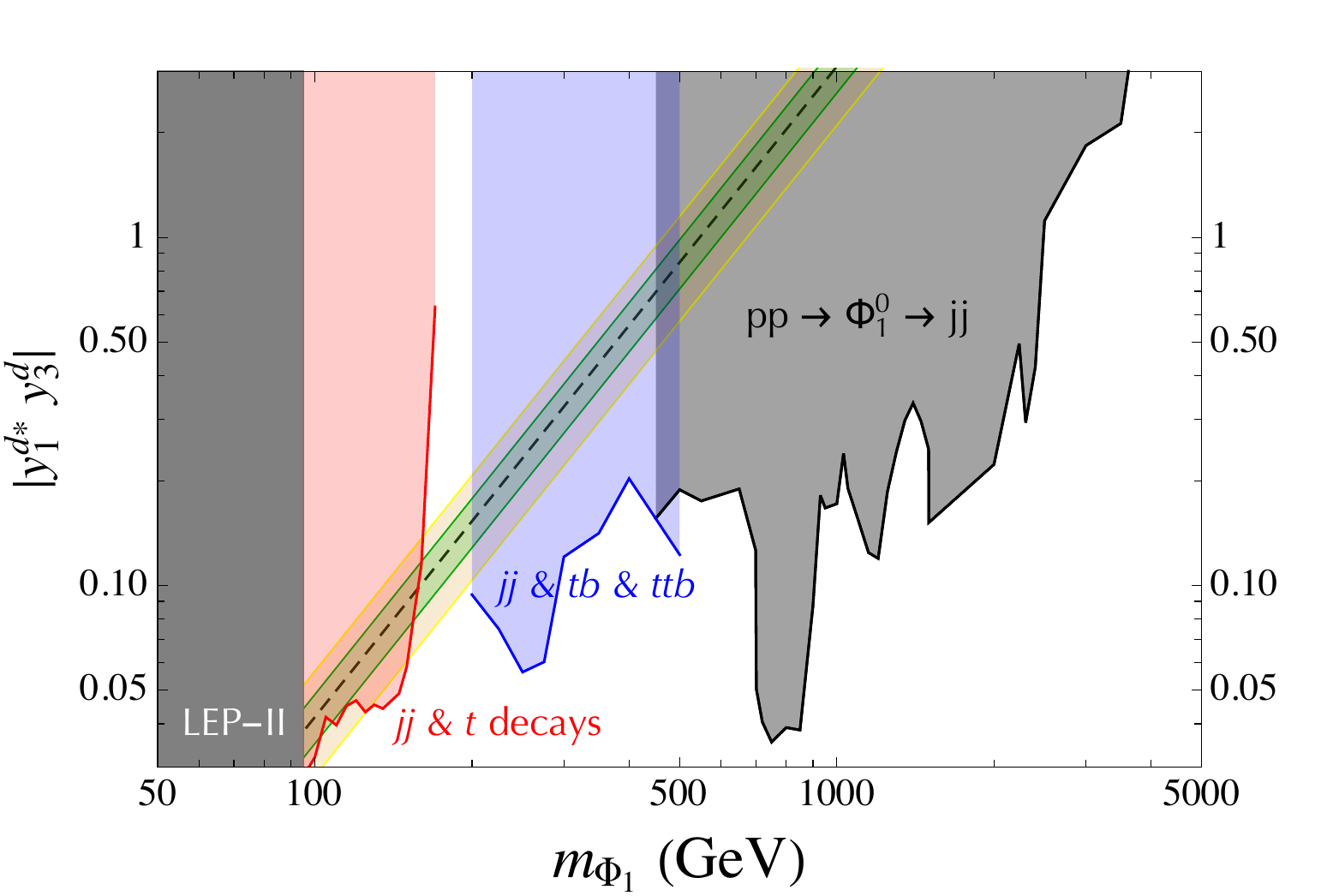} 
\caption{\label{fig:Higgs} The compilation of the high-$p_T$ collider constraints on the $\Phi_1$ model (Benchmark I) together with the best-fit region from non-leptonic $B$ decays. See Section~\ref{sec:Higgs} for details.}
\end{figure}

Despite different flavor structures, the two benchmarks share a similar collider phenomenology. We focus on one of them for simplicity. The leading high-$p_T$ constraints on the first benchmark are complied in Fig.~\ref{fig:Higgs} for the entire valid range of $m_{\Phi_1}$. The limits from single dijet resonance searches, compared with the region preferred by the non-leptonic $B$ decays, firmly exclude masses above $m_{\Phi_1} \gtrsim 450$ GeV as the explanation of the anomaly.
These are obtained by applying the general results from Fig.~\ref{fig:dijetNC} using Eq.~\eqref{eq:inequality} for the charge-neutral component of the $SU(2)_L$ doublet. Note that this  component exclusively decays to a dijet final state for all $m_{\Phi_1}$, unlike the charged one, which has a significant branching ratio to a top quark and a jet when $m_{\Phi_1} > m_t$. The introduction of large mass splitting between the two $SU(2)_L$ components potentially helps to avoid bounds on the neutral component.  However, this is very well constrained by the electroweak precision tests, see for example~\cite{Faroughy:2016osc}.
While the single dijet searches at the LHC are as effective as for other mediators, on the contrary, we find the pair-produced dijet searches do not probe the leftover parameter space of interest. The reason for this is that the pair production cross section for uncolored particles at the LHC is not large enough to match the current sensitivity. The lower limit on the mass comes only from LEP-II as discussed in Section~\ref{sec:highpTpair}, in particular, $m_{\Phi_1} \gtrsim 95$~GeV.

We are left with the mass range $m_{\Phi_1} \in [95-450]$~GeV. To attack this parameter space, let us consider the CMS search for a light dijet resonance reported in Ref.~\cite{Sirunyan:2019vxa}. The resonance has a large $p_T$ as it recoils from another jet in the production, i.e. $p p \to X j$ process. The $X$ decay products are collimated leading to a fat jet topology. The signal model used in this analysis is a leptophobic $Z'$ with flavor universal vector-like couplings to quarks. The collaboration performed a simulation of $Z'+$jet(s) with the parton-level filter $H_T  > 400$~GeV (for details see~\cite{Sirunyan:2019vxa}). The events are selected by asking for at least one AK8 jet (anti-$k_T$ with $R=0.8$) with $p_T > 525$\,GeV for $m_X < 175$\,GeV or CA15 jet (Cambridge-Aachen with $R=1.5$) with $p_T > 575$\,GeV for $m_X > 175$\,GeV. In both cases, the pseudorapidity is $|\eta| < 2.5$. The exclusion limits are reported on the coupling $g'_q$ as a function of the $Z'$ mass, see Fig.~4 in~\cite{Sirunyan:2019vxa}. To translate these limits to our case, we perform the {\tt MadGraph5\_aMC@NLO}~\cite{Alwall:2014hca} partonic level simulation of $p p \to H' j \to (j j) j$ imposing the same $H_T$, $\eta$, and $p_T$ cuts, as well as, $\Delta R < 0.8 \, (1.5)$ for the resonance decay products. We repeat the same procedure for the $Z'$ model used by the CMS collaboration. The translation is done by comparing the two fiducial cross sections for several benchmark masses.\footnote{This procedure will be followed to derive bounds for mediators with different spins and representations in the future study. Also, the partonic-level simulation gives only an estimate of the difference in the signal acceptance between $H'$ and $Z'$ models and will be superseded by a full-fledged detector-level simulation.}

In addition to this, important constraints come from the top sector. In particular, searches for $\Phi_1 \to t b$, when $m_{\Phi_1} > m_t$,  and for $t \to b \Phi_1$, when $m_{\Phi_1} < m_t$, exclude most of the remaining parameter space except for a small window around $m_{\Phi_1} \approx m_t$, where the two processes are kinematically suppressed (see Fig.~\ref{fig:Higgs}). The  bounds on $y_3^d$ from top decays are  extracted from the dedicated ATLAS search~\cite{Sirunyan:2020aln}. The exclusion shown in red color in Fig.~\ref{fig:Higgs} is the product of this bound with the dijet bound from~\cite{Sirunyan:2019vxa}. Similarly, the exclusion limits shown in blue color are a combination of the ATLAS search in the $t \bar t b$ final state~\cite{ATLAS:2020jqj}, the CDF search for a resonance in $p p \to t b$~\cite{Aaltonen:2009qu,Bianchi:2011xsw} and the dijet search~\cite{Sirunyan:2019vxa}.

In conclusion, there is a blind spot around $m_{\Phi_1} \approx m_t $ which needs a dedicated collider search to be covered. We used {\tt MadGraph5\_aMC@NLO}~\cite{Alwall:2014hca} to calculate processes with off-shell $\Phi_1$ and/or top quark. We identify the following signatures at the LHC which could further squeeze the interesting parameter space: $t \bar t \Phi_1$, single top, and $V\Phi_1$ where $V=W,Z$. Further improvements of the light dijet resonance searches such as~\cite{Sirunyan:2019vxa} would also be beneficial.

\subsubsection*{Flavour Constraints}

Concerning constraints from low-energy observables, we note that in Benchmark II flavor changing processes from a $b$ to lighter down-type quarks are always proportional to $y^u_1$ and are then suppressed by the up-quark mass. Similar arguments can be applied to charm physics, where only the coupling $y^d_3$ enters. However, in this case, the strong suppression comes from the CKM. Therefore we conclude that Benchmark II is insensitive to flavor constraints.
In the case of Benchmark I, it is not straightforward to draw analogous conclusions. In \app{sec:HiggsFlavor} we investigate a specific parameter-space point for $m_{\phi_1} \approx m_t$ and $y^d_i$, finding that also Benchmark I cannot be excluded by low-energy flavour constraints. Small variations around this point do not change our conclusions.

\section{Conclusions}
\label{sec:Conc}

The size of potential new physics effects in non-leptonic $B$ meson decays is restricted by complementary processes. A typical example is neutral meson mixing, which is predicted to accompany non-leptonic decays in many extensions of the SM. $\Delta F = 2$ transitions are usually the leading constraints on such models, implying unobservable effects in non-leptonic decays. It is, however, not difficult to find exceptions to this prevalent expectation. In this work, we provide examples of untuned models owning specific flavor structure and dynamical content, that are safe on all complementary processes in flavor physics. Nevertheless, we also find that such cases are severely constrained by the dijet resonance searches at high-$p_T$ colliders. Intuitively, new mediators have to compete with the tree-level $W$ boson exchange, and to have an appreciable effect, their masses and couplings turned out to be perfectly suited for high-$p_T$ searches at the LHC. The connection is even more profound since the minimal set of couplings needed to generate the desired effect in non-leptonic decays can not be neglected in high-$p_T$ dijet searches, unlike for example in $\Delta F = 2$ where the effect can be loop-induced and flavor-suppressed.

Recognising the importance of dijet searches for non-leptonic decays, we reinterpret the existing ATLAS and CMS searches as a constraint on a generic narrow spin-0 and spin-1 resonance, which can be a color singlet, triplet, sextet, or octet, coupled to a diquark or a quark-antiquark pair with an arbitrary flavor composition. The pair production of colored resonances $p p \to X X \to (jj) (jj)$ decaying dominantly to jets sets a robust lower limit on their mass as explained in \sec{sec:highpTpair}. The single dijet resonance production $p p \to X \to jj$ is discussed in \sec{sec:highpTsingle} and the main results are summarised in Figs.~\ref{fig:dijet} and \ref{fig:dijetNC}. Since different flavor channels do not interfere, the absence of the signal imposes limits on all of them simultaneously. The results derived in this section go beyond $\bar B_q\to D_q^{(*)+}  \,\{\pi, K\}$ decays and can be used to limit new physics contributions in other hadronic decays.

Regarding $\bar B_q\to D_q^{(*)+}  \,\{\pi, K\}$ decays, we perform a full EFT analysis for arbitrary new physics at leading-order in $\alpha_s$, see \sec{sec:EFT}. The main results are shown in Fig.~\ref{fig:1}. The discrepancy between the SM theory prediction and measurements observed in Ref.~\cite{Bordone:2020gao} leads to the best-fit region favoring new physics. We find this as an excellent opportunity to test the importance of dijet constraints. Based on the EFT analysis, we write an exhaustive list of simplified mediator models which can be matched to the best-fit region at low energies. These are reported in Eq.~\eqref{eq:mediators}.

Assuming pragmatically the minimal set of couplings needed to fit the anomaly, we study the collider limits for every viable {tree-level} mediator. For colored mediators, in \sec{sec:coloredM}, we show that the combination of pair and single production of dijet resonances excludes weakly-coupled new physics explanations of the anomalies, as illustrated in Fig.~\ref{fig:colored_vs_dijet}. The pair production is effective for light resonances, while the single production excludes heavier resonances.
Besides, we also carry out a thorough study of flavor phenomenology for the color-sextet scalar model to show that other low-energy bounds such as $\Delta F =2$ can naturally be avoided, see \app{sec:SextetFlavor}. Thus, the high-$p_T$ dijet constraints are crucial to dismiss this model as the origin of the discrepancy. 

Finally, we study the colorless weak-doublet scalar $\Phi_1$ in \sec{sec:Higgs} which is not sufficiently constrained by the pair production at the LHC. However, we show that the consistency of flavor data requires a specific flavor alignment, predicting sizeable couplings with the top quark. As a consequence, top-quark physics rules out the interesting parameter space for all hypothetical $m_{\Phi_1}$ apart from a narrow window around the top mass, see Fig.~\ref{fig:Higgs}. This blind point is quite resilient on flavor bounds which are carefully checked in \app{sec:HiggsFlavor}, and will likely be covered by the future collider studies involving the top quark, as well as by  further progress in dijet resonance searches.

\acknowledgments
	
	We thank Javier Fuentes-Mart\'{\i}n for collaboration in the early stages of this work. We also thank Martin Jung for useful communications. The work of MB is supported by Deutsche Forschungsgemeinschaft (DFG, German Research Foundation) under grant 396021762 - TRR 257 ``Particle Physics Phenomenology after the Higgs Discovery'' and by the Italian Ministry of Research (MIUR) under grant PRIN 20172LNEEZ. DM acknowledges support by the INFN grant SESAMO and MIUR grant PRIN 2017L5W2PT. The work of AG has received funding from the Swiss National Science Foundation (SNF) through the Eccellenza Professorial Fellowship ``Flavor Physics at the High Energy Frontier'' project number 186866. This work is also partially supported by the European Research Council (ERC) under the European Union’s Horizon 2020 research and innovation programme, grant agreement 833280 (FLAY).

\appendix
	
\section{Details of the EFT analysis}

\subsection*{Hadronic matrix elements for the NP operators}
\label{app:first}
The set of operators in the LEFT that we consider is:
\begin{align}
[\mathcal{O}_{ud}^{V1,LL}]_{ijkl}&=(\bar u_L^i\gamma_\mu u_L^j)(\bar d_L^k\gamma^\mu d_L^l)  \,, &  
[\mathcal{O}_{ud}^{V8,LL}]_{ijkl}&=(\bar u_L^i\gamma_\mu T^A u_L^j)		(\bar d_L^k\gamma^\mu  T^A d_L^l)  \,, \nonumber\\
[\mathcal{O}_{ud}^{V1,RR}]_{ijkl}&=(\bar u_R^i\gamma_\mu u_R^j)(\bar d_R^k\gamma^\mu d_R^l)  \,, & 
[\mathcal{O}_{ud}^{V8,RR}]_{ijkl}&=(\bar u_R^i\gamma_\mu T^A u_R^j)		(\bar d_R^k\gamma^\mu  T^A d_R^l)  \,, \nonumber\\
[\mathcal{O}_{ud}^{V1,LR}]_{ijkl}&=(\bar u_L^i\gamma_\mu u_L^j)(\bar d_R^k\gamma^\mu d_R^l) \,, &  
[\mathcal{O}_{ud}^{V8,LR}]_{ijkl}&=(\bar u_L^i\gamma_\mu T^A u_L^j)(\bar d_R^k\gamma^\mu T^A d_R^l) \,, \nonumber\\
[\mathcal{O}_{du}^{V1,LR}]_{ijkl}&=(\bar d_L^i\gamma_\mu d_L^j) (\bar u_R^k\gamma^\mu u_R^l)  \,, &  
[\mathcal{O}_{du}^{V8,LR}]_{ijkl}&=(\bar d_L^i\gamma_\mu T^A d_L^j)(\bar u_R^k\gamma^\mu T^A u_R^l)   \,, \nonumber\\
[\mathcal{O}_{uddu}^{V1,LR}]_{ijkl}&=(\bar{u}^i_L\gamma_\mu d_L^j)(\bar{d}^k_R \gamma^\mu u_R^l)\,   \,, &  
[\mathcal{O}_{uddu}^{V8,LR}]_{ijkl}&=(\bar{u}^i_L\gamma_\mu T^A  d_L^j)(\bar{d}^k_R \gamma^\mu  T^A u_R^l)\,     \,, \nonumber\\
[\mathcal{O}_{ud}^{S1,RR}]_{ijkl}&=(\bar{u}_L^i u_R^j)(\bar{d}_L^k d_R^l)  \,, &  
[\mathcal{O}_{ud}^{S8,RR}]_{ijkl}&=(\bar{u}_L^i T^A u_R^j)(\bar{d}_L^k T^A d_R^l)  \,, \nonumber\\
[\mathcal{O}_{uddu}^{S1,RR}]_{ijkl}&=(\bar{u}_L^i d_R^j)(\bar{d}_L^k u_R^l)    \,, &  
[\mathcal{O}_{uddu}^{S8,RR}]_{ijkl}&=(\bar{u}_L^i T^A d_R^j)(\bar{d}_L^k T^A u_R^l)  
\label{eq:LEFT_basis}
\end{align}
In order to evaluate the necessary matrix elements, apply Fierz transformation to get the following set of operators:
\begin{align}
\mathcal{Q}_{V_{LL}}^{\prime ijkl}&=(\bar u_L^i \gamma_\mu T^A d_L^j)(\bar d_L^k \gamma^\mu T^A u_L^l) \,,  &
\mathcal{Q}_{V_{LL}}^{ijkl}&=(\bar u_L^i \gamma_\mu d_L^j)(\bar d_L^k \gamma^\mu u_L^l) \,,\nonumber\\
\mathcal{Q}_{V_{RR}}^{\prime ijkl}&=(\bar u _R^i \gamma_\mu T^A d_R^j)(\bar d_R^k \gamma^\mu T^A u_R^l)\,,&
\mathcal{Q}_{V_{RR}}^{ijkl}&=(\bar u_R^i \gamma_\mu d_R^j)(\bar d_R^k \gamma^\mu u_R^l) \,,\nonumber\\
\mathcal{Q}_{V_{LR}}^{\prime ijkl}&=(\bar u_L^i\gamma_\mu T^A  d_L^j)(\bar d_R^k \gamma^\mu  T^A u_R^l) \,,&
\mathcal{Q}_{V_{LR}}^{ijkl}&=(\bar u_L^i\gamma_\mu d_L^j)(\bar d_R^k \gamma^\mu u_R^l) \,,\nonumber\\
\mathcal{Q}_{S_{RL}}^{\prime ijkl}&=(\bar u_L^i T^A  d_R^j)(\bar d_R^k T^A  u_L^l)\,,&
\mathcal{Q}_{S_{RL}}^{ijkl}&=(\bar u_L^i  d_R^j)(\bar d_R^k u_L^l)\,,\nonumber\\
\mathcal{Q}_{S_{LR}}^{\prime ijkl}&=(\bar u_R^i T^A  d_L^j)(\bar d_L^k  T^A u_R^l)\,,&
\mathcal{Q}_{S_{LR}}^{ijkl}&=(\bar u_R^i   d_L^j)(\bar d_L^k  u_R^l)\nonumber\,,\\
\mathcal{Q}_{S_{RR}}^{\prime ijkl}&=(\bar u_L^i T^A  d_R^j)(\bar d_L^k  T^A u_R^l) \,,&
\mathcal{Q}_{S_{RR}}^{ijkl}&=(\bar u_L^i   d_R^j)(\bar d_L^k  u_R^l)  \,,\nonumber\\
\mathcal{Q}_{T_{RR}}^{\prime ijkl}&=(\bar u_L^i \sigma_{\mu\nu} T^A  d_R^j)(\bar d_L^k  \sigma^{\mu\nu} T^A u_R^l) \,,&
\mathcal{Q}_{T_{RR}}^{ijkl}&=(\bar u_L^i \sigma_{\mu\nu}  d_R^j)(\bar d_L^k  \sigma^{\mu\nu} u_R^l)  \,,
\label{eq:basis_Q}
\end{align}
The relation between the two sets of operators is given by 
\begin{align}
a_{V_{LL}}^{\prime ijkl}&=2\,[L_{ud}^{V1,LL}]_{ilkj}-\frac{1}{3}\,[L_{ud}^{V1,LL}]_{ilkj}\,,  &
a_{V_{LL}}^{ijkl}&=\frac{1}{3}\,[L_{ud}^{V1,LL}]_{ilkj}+\frac{4}{9}\,[L_{ud}^{V1,LL}]_{ilkj}\,,\nonumber\\
a_{V_{RR}}^{\prime ijkl}&=2\,[L_{ud}^{V1,RR}]_{ilkj}-\frac{1}{3}\,[L_{ud}^{V1,RR}]_{ilkj}\,,&
a_{V_{RR}}^{ijkl}&=\frac{1}{3}\,[L_{ud}^{V1,RR}]_{ilkj}+\frac{4}{9}\,[L_{ud}^{V1,RR}]_{ilkj}\,,\nonumber\\
a_{V_{LR}}^{\prime ijkl}&=[L_{uddu}^{V1,LR}]_{ijkl}\,,&
a_{V_{LR}}^{ijkl}&=[L_{uddu}^{V1,LR}]_{ijkl}\,,\nonumber\\
a_{S_{RL}}^{\prime ijkl}&=-4\,[L_{ud}^{V1,LR}]_{ilkj}+\frac{2}{3}\,[L_{ud}^{V1,LR}]_{ilkj}\,,&
a_{S_{RL}}^{ijkl}&=-\frac{2}{3}\,[L_{ud}^{V1,LR}]_{ilkj}-\frac{8}{9}\,[L_{ud}^{V1,LR}]_{ilkj}\,,\nonumber\\
a_{S_{LR}}^{\prime ijkl}&=-4\,[L_{du}^{V1,LR}]_{kjil}+\frac{2}{3}\,[L_{du}^{V1,LR}]_{kjil}\,,&
a_{S_{LR}}^{ijkl}&=-\frac{2}{3}\,[L_{du}^{V1,LR}]_{kjil}-\frac{8}{9}\,[L_{du}^{V1,LR}]_{kjil}\nonumber\,,\\
a_{S_{RR}}^{\prime ijkl}&=-[L_{ud}^{S1,RR}]_{ilkj}+\frac{1}{6}\,[L_{ud}^{S8,RR}]_{ilkj}+[L_{uddu}^{S8,RR}]_{ijkl}\,,& \nonumber\\
a_{S_{RR}}^{ijkl}&=-\frac{1}{6}\,[L_{ud}^{S1,RR}]_{ilkj}-\frac{2}{9}\,[L_{ud}^{S8,RR}]_{ilkj}+[L_{uddu}^{S1,RR}]_{ijkl}\,, & \nonumber\\
a_{T_{RR}}^{\prime ijkl}&=-\frac{1}{4}\,[L_{ud}^{S1,RR}]_{ilkj}+\frac{1}{24}\,[L_{ud}^{S8,RR}]_{ilkj}\,,&
a_{T_{RR}}^{ijkl}&=-\frac{1}{24}\,[L_{ud}^{S1,RR}]_{ilkj}-\frac{1}{18}\,[L_{ud}^{S8,RR}]_{ilkj}\,
\label{eq:matching_LEFT_Q}
\end{align}
We can now evaluate the hadronic matrix elements for the operators $\mathcal{Q}_i$ at leading power in $1/m_b$ and leading order in $\alpha_s$. We note that, due to color algebra, the matrix elements  $\bra{P^- D_q^{+}}\mathcal{Q}_i^\prime\ket{\bar B_q} =0$ at leading order in $\alpha_s$. We further notice that $\bra{P^- D_q^{+}}\mathcal{Q}^{(\prime)}_{T_{RR}}\ket{\bar B_q} =0 $ at any order.\\
For a pseudoscalar $D_q^+$ in the final state we obtain:
\begin{align}
\bra{P^- D_q^{+}}\mathcal{Q}_{V_{LL}}^{cbiu}\ket{\bar B_q} &= -\frac{i}{4} f_P (\mBq^2-\mDqst^2) F_0^{\bar B_q\to D_q}\,,  \nonumber\\
\bra{P^- D_q^{+}}\mathcal{Q}_{V_{RR}}^{cbiu}\ket{\bar B_q} &=+\frac{i}{4} f_P (\mBq^2-\mDq^2) F_0^{\bar B_q\to D_q}\,, \nonumber\\
\bra{P^- D_q^{+}}\mathcal{Q}_{V_{LR}}^{cbiu}\ket{\bar B_q} &= +\frac{i}{4} f_P (\mBq^2-\mDq^2) F_0^{\bar B_q\to D_q}\,, \nonumber\\
\bra{P^- D_q^{+}}\mathcal{Q}_{V_{LR}}^{uibc}\ket{\bar B_q} &=  -\frac{i}{4} f_P (\mBq^2-\mDq^2) F_0^{\bar B_q\to D_q}\,, \nonumber\\
\bra{P^- D_q^{+}}\mathcal{Q}_{S_{RL}}^{cbiu}\ket{\bar B_q} &=+\frac{i}{4} f_P (\mBq^2-\mDq^2) F_0^{\bar B_q\to D_q} \frac{m_P^2}{(m_u+m_{d_i})(m_b-m_c)}\,,\nonumber\\
\bra{P^- D_q^{+}}\mathcal{Q}_{S_{LR}}^{cbiu}\ket{\bar B_q} &= -\frac{i}{4} f_P (\mBq^2-\mDq^2) F_0^{\bar B_q\to D_q} \frac{m_P^2}{(m_u+m_{d_i})(m_b-m_c)}\,,\nonumber\\
\bra{P^- D_q^{+}}\mathcal{Q}_{S_{RR}}^{cbiu}\ket{\bar B_q} &=-\frac{i}{4} f_P (\mBq^2-\mDq^2) F_0^{\bar B_q\to D_q} \frac{m_P^2}{(m_u+m_{d_i})(m_b-m_c)}\,,\nonumber\\
\bra{P^- D_q^{+}}\mathcal{Q}_{S_{RR}}^{uibc}\ket{\bar B_q} &=+\frac{i}{4} f_P (\mBq^2-\mDq^2) F_0^{\bar B_q\to D_q} \frac{m_P^2}{(m_u+m_{d_i})(m_b-m_c)}\,,\nonumber\\
\bra{P^- D_q^{+}}\mathcal{Q}_{T_{RR}}^{cbiu}\ket{\bar B_q} &=\,0\,,\nonumber\\
\bra{P^- D_q^{+}}\mathcal{Q}_{T_{RR}}\ket{\bar B_q} &=\,0\,,
\label{eq:had_D}
\end{align}
where $f_P$ is the decay constant of the $P^-$ meson and $F_0$ is the scalar form factor for $\bar{B}_q\to D_q^+$ decay as in Ref.~\cite{Bordone:2019guc}. For a vector $D_q^{*+}$ in the final state we obtain:
 \begin{align}
\bra{P^- D_q^{*+}}\mathcal{Q}_{V_{LL}}^{cbiu}\ket{\bar B_q}  &= +\frac{i}{4} f_P \sqrt{\lambda_P} \,A_0^{\bar B_q\to D^*_q} \,, \nonumber \\
\bra{P^- D_q^{*+}}\mathcal{Q}_{V_{RR}}^{cbiu}\ket{\bar B_q} &= +\frac{i}{4}f_P \sqrt{\lambda_P} \,A_0^{\bar B_q\to D^*_q} \,,\nonumber \\
\bra{P^- D_q^{*+}}\mathcal{Q}_{V_{LR}}^{cbiu}\ket{\bar B_q} &=  -\frac{i}{4} f_P \sqrt{\lambda_P}\, A_0^{\bar B_q\to D^*_q}\,,\nonumber \\
\bra{P^- D_q^{*+}}\mathcal{Q}_{V_{LR}}^{uibc}\ket{\bar B_q} &= -\frac{i}{4} f_P \sqrt{\lambda_P}\, A_0^{\bar B_q\to D^*_q}\,,\nonumber \\
\bra{P^- D_q^{*+}}\mathcal{Q}_{S_{RL}}^{cbiu}\ket{\bar B_q} &=-\frac{i}{4} f_P \sqrt{\lambda_P}\, A_0^{\bar B_q\to D^*_q} \frac{m_P^2}{(m_u+m_{d_i})(m_b+m_c)} \,,\nonumber\\
\bra{P^- D_q^{*+}}\mathcal{Q}_{S_{LR}}^{cbiu}\ket{\bar B_q} &= -\frac{i}{4} f_P \sqrt{\lambda_P} \,A_0^{\bar B_q\to D^*_q} \frac{m_P^2}{(m_u+m_{d_i})(m_b+m_c)} \,,\nonumber\\
\bra{P^- D_q^{*+}}\mathcal{Q}_{S_{RR}}^{cbiu}\ket{\bar B_q} &=+\frac{i}{4} f_P \sqrt{\lambda_P} \,A_0^{\bar B_q\to D^*_q} \frac{m_P^2}{(m_u+m_{d_i})(m_b+m_c)}\,,\nonumber\\
\bra{P^- D_q^{*+}}\mathcal{Q}_{S_{RR}}^{uibc}\ket{\bar B_q} &=+\frac{i}{4} f_P \sqrt{\lambda_P} \,A_0^{\bar B_q\to D^*_q} \frac{m_P^2}{(m_u+m_{d_i})(m_b+m_c)}\,,\nonumber\\
\bra{P^- D_q^{*+}}\mathcal{Q}_{T_{RR}}^{cbiu}\ket{\bar B_q} &=\, 0\,,\nonumber\\
\bra{P^- D_q^{*+}}\mathcal{Q}_{T_{RR}}^{uibc}\ket{\bar B_q}&=\, 0\,,
\label{eq:had_Dst}
\end{align}
where $A_0$ is the scalar form factor for $\bar{B}_q\to D_q^{+*}$ decay as in Ref.~\cite{Bordone:2019guc}, and $\lambda_P = \mBq^4+\mDqst^4+m_P^4 - 2m_P^2 (\mBq^2+\mDqst^2)-2 \mBq^2 \mDqst^2$.

\subsection*{LEFT RGE}
\label{app:LEFTRGE}

Here we give the one-loop renormalization group equations for the LEFT coefficients.
We run from the bottom mass scale up to the EW scale with 5 quark flavors.
The solutions to the one-loop RGE is $\vec{L}(m_b) = U(m_b, m_{Z}) \vec{L}(m_{Z})$, where the vector of EFT coefficients is 
\be
\vec{L} =  \left(
\begin{array}{c} 
	{[} L_{ud}^{V1,LL} {]}_{21i3} \\
	{[} L_{ud}^{V8,LL} {]}_{21i3} \\
	{[} L_{ud}^{V1,RR} {]}_{21i3} \\
	{[} L_{ud}^{V8,RR} {]}_{21i3} \\
	{[} L_{uddu}^{V1,LR} {]}_{23i1} \\
	{[} L_{uddu}^{V8,LR} {]}_{23i1} \\
	{[} L_{uddu}^{V1,LR} {]}_{1i32} \\
	{[} L_{uddu}^{V8,LR} {]}_{1i32} \\
	{[} L_{ud}^{V1,LR} {]}_{21i3} \\
	{[} L_{ud}^{V8,LR} {]}_{21i3} \\
	{[} L_{du}^{V1,LR} {]}_{i321} \\
	{[} L_{du}^{V8,LR} {]}_{i321} \\
	{[} L_{uddu}^{S1,RR} {]}_{23i1} \\
	{[} L_{uddu}^{S8,RR} {]}_{23i1} \\
	{[} L_{ud}^{S1,RR} {]}_{21i3} \\
	{[} L_{ud}^{S8,RR} {]}_{21i3} \\
	{[} L_{uddu}^{S1,RR} {]}_{1i32} \\
	{[} L_{uddu}^{S8,RR} {]}_{1i32} \\
	{[} L_{ud}^{S1,RR} {]}_{123i} \\
	{[} L_{ud}^{S8,RR} {]}_{123i} 
\end{array}
\right)
\ee
and, using one-loop anomalous dimensions of LEFT operators \cite{Jenkins:2017dyc} implemented in DsixTools 2.0 \cite{Celis:2017hod,Fuentes-Martin:2020zaz} we get
\be\begin{split}
\hspace{-1cm} 
&U(m_b, m_{Z}) = \\
&{\tiny\left(
\begin{array}{cccccccccccccccccccc}
 1.03 & -0.1 & 0 & 0 & 0 & 0 & 0 & 0 & 0 & 0 & 0 & 0 & 0 & 0 & 0 & 0 & 0 & 0 & 0 & 0 \\
 -0.47 & 1.18 & 0 & 0 & 0 & 0 & 0 & 0 & 0 & 0 & 0 & 0 & 0 & 0 & 0 & 0 & 0 & 0 & 0 & 0 \\
 0 & 0 & 1.03 & -0.1 & 0 & 0 & 0 & 0 & 0 & 0 & 0 & 0 & 0 & 0 & 0 & 0 & 0 & 0 & 0 & 0 \\
 0 & 0 & -0.47 & 1.18 & 0 & 0 & 0 & 0 & 0 & 0 & 0 & 0 & 0 & 0 & 0 & 0 & 0 & 0 & 0 & 0 \\
 0 & 0 & 0 & 0 & 1.03 & 0.13 & 0 & 0 & 0 & 0 & 0 & 0 & 0 & 0 & 0 & 0 & 0 & 0 & 0 & 0 \\
 0 & 0 & 0 & 0 & 0.57 & 1.7 & 0 & 0 & 0 & 0 & 0 & 0 & 0 & 0 & 0 & 0 & 0 & 0 & 0 & 0 \\
 0 & 0 & 0 & 0 & 0 & 0 & 1.03 & 0.13 & 0 & 0 & 0 & 0 & 0 & 0 & 0 & 0 & 0 & 0 & 0 & 0 \\
 0 & 0 & 0 & 0 & 0 & 0 & 0.57 & 1.7 & 0 & 0 & 0 & 0 & 0 & 0 & 0 & 0 & 0 & 0 & 0 & 0 \\
 0 & 0 & 0 & 0 & 0 & 0 & 0 & 0 & 1.02 & 0.13 & 0 & 0 & 0 & 0 & 0 & 0 & 0 & 0 & 0 & 0 \\
 0 & 0 & 0 & 0 & 0 & 0 & 0 & 0 & 0.57 & 1.68 & 0 & 0 & 0 & 0 & 0 & 0 & 0 & 0 & 0 & 0 \\
 0 & 0 & 0 & 0 & 0 & 0 & 0 & 0 & 0 & 0 & 1.02 & 0.13 & 0 & 0 & 0 & 0 & 0 & 0 & 0 & 0 \\
 0 & 0 & 0 & 0 & 0 & 0 & 0 & 0 & 0 & 0 & 0.57 & 1.68 & 0 & 0 & 0 & 0 & 0 & 0 & 0 & 0 \\
 0 & 0 & 0 & 0 & 0 & 0 & 0 & 0 & 0 & 0 & 0 & 0 & 1.82 & -0.04 & -0.44 & -0.17 & 0 & 0 & 0 & 0.
   \\
 0 & 0 & 0 & 0 & 0 & 0 & 0 & 0 & 0 & 0 & 0 & 0 & -0.42 & 0.82 & 0.34 & -0.13 & 0 & 0 & 0 & 0 \\
 0 & 0 & 0 & 0 & 0 & 0 & 0 & 0 & 0 & 0 & 0 & 0 & -0.44 & -0.17 & 1.84 & -0.04 & 0 & 0 & 0 & 0.
   \\
 0 & 0 & 0 & 0 & 0 & 0 & 0 & 0 & 0 & 0 & 0 & 0 & 0.32 & -0.13 & -0.42 & 0.84 & 0 & 0 & 0 & 0 \\
 0 & 0 & 0 & 0 & 0 & 0 & 0 & 0 & 0 & 0 & 0 & 0 & 0 & 0 & 0 & 0 & 1.82 & -0.04 & -0.44 & -0.17
   \\
 0 & 0 & 0 & 0 & 0 & 0 & 0 & 0 & 0 & 0 & 0 & 0 & 0 & 0 & 0 & 0 & -0.42 & 0.82 & 0.34 & -0.13 \\
 0 & 0 & 0 & 0 & 0 & 0 & 0 & 0 & 0 & 0 & 0 & 0 & 0 & 0 & 0 & 0 & -0.44 & -0.17 & 1.84 & -0.04
   \\
 0 & 0 & 0 & 0 & 0 & 0 & 0 & 0 & 0 & 0 & 0 & 0 & 0 & 0 & 0 & 0 & 0.32 & -0.13 & -0.42 & 0.84 \\
\end{array}
\right)
}
\end{split}\ee
%

\subsection*{SMEFT operators and matching to LEFT}
\label{sec:SMEFT_ops}

The SMEFT operators relevant to this process are
\be
	\mathcal{L}_{\rm SMEFT} = \sum_i C_i \mathcal{O}_i~,
\ee
with
\begin{align}
[\mathcal{O}_{qq}^{(1)}]_{ijkl}&= (\bar q_L^i \gamma_\mu q_L^j)(\bar q_L^k \gamma_\mu q_L^l)  \,, &  
[\mathcal{O}_{qq}^{(3)}]_{ijkl}&= (\bar q_L^i \sigma^a \gamma_\mu q_L^j)(\bar q_L^k \sigma^a \gamma_\mu q_L^l) \,, \nonumber\\
[\mathcal{O}_{ud}^{(1)}]_{ijkl}&= (\bar u_R^i \gamma_\mu u_R^j)(\bar d_R^k \gamma_\mu d_R^l)  \,, &  
[\mathcal{O}_{ud}^{(8)}]_{ijkl}&= (\bar u_R^i T^A \gamma_\mu u_R^j)(\bar d_R^k T^A \gamma_\mu d_R^l) \,, \nonumber\\
[\mathcal{O}_{qd}^{(1)}]_{ijkl}&=  (\bar q_L^i \gamma_\mu q_L^j)(\bar d_R^k \gamma_\mu d_R^l)  \,, &  
[\mathcal{O}_{qd}^{(8)}]_{ijkl}&=  (\bar q_L^i T^A \gamma_\mu q_L^j)(\bar d_R^k T^A \gamma_\mu d_R^l) \,, \nonumber\\
[\mathcal{O}_{qu}^{(1)}]_{ijkl}&=  (\bar q_L^i \gamma_\mu q_L^j)(\bar u_R^k \gamma_\mu u_R^l)  \,, &  
[\mathcal{O}_{qu}^{(8)}]_{ijkl}&=  (\bar q_L^i T^A \gamma_\mu q_L^j)(\bar u_R^k T^A \gamma_\mu u_R^l) \,, \nonumber\\
[\mathcal{O}_{quqd}^{(1)}]_{ijkl}&=  (\bar q_L^i u_R^j) (i\sigma^2) (\bar q_L^k d_R^l)  \,, &  
[\mathcal{O}_{quqd}^{(8)}]_{ijkl}&=  (\bar q_L^i T^A u_R^j) (i\sigma^2) (\bar q_L^k T^A  d_R^l) \,. \nonumber\\
\label{eq:SMEFT_basis}
\end{align}
Tree-level matching between the LEFT and the SMEFT, in the down-quark mass basis:
\begin{align}
	[L_{ud}^{V1,LL}]_{prst} &= V_{pi} V^*_{rj} \left( [C_{qq}^{(1)}]_{ijst} + [C_{qq}^{(1)}]_{stij} - [C_{qq}^{(3)}]_{ijst} - [C_{qq}^{(3)}]_{stij} +\frac{2}{N_c} ( [C_{qq}^{(3)}]_{itsj} + [C_{qq}^{(3)}]_{sjit} ) \right) \nonumber\\
	[L_{ud}^{V8,LL}]_{prst} &= 4 V_{pi} V^*_{rj} \left( [C_{qq}^{(3)}]_{itsj} + [C_{qq}^{(3)}]_{sjit} \right)  \nonumber\\
	[L_{ud}^{V1,RR}]_{prst} &=  [C_{ud}^{(1)}]_{prst} \nonumber\\
	[L_{ud}^{V8,RR}]_{prst} &=  [C_{ud}^{(8)}]_{prst} \nonumber\\
	[L_{ud}^{V1,LR}]_{prst} &= 4 V_{pi} V^*_{rj} [C_{qd}^{(1)}]_{ijst}\nonumber \\
	[L_{ud}^{V8,LR}]_{prst} &= 4 V_{pi} V^*_{rj} [C_{qd}^{(8)}]_{ijst} \nonumber\\
	[L_{du}^{V1,LR}]_{prst} &=  [C_{qu}^{(1)}]_{prst} \nonumber\\
	[L_{du}^{V8,LR}]_{prst} &=  [C_{qu}^{(8)}]_{prst} \label{eq:LEFtoSMEFT}\\
	[L_{uddu}^{V1,LR}]_{prst} &= 0 \nonumber\\
	[L_{uddu}^{V8,LR}]_{prst} &= 0 \nonumber\\
	[L_{ud}^{S1,RR}]_{prst} &= V_{pi} [C_{quqd}^{(1)}]_{irst} \nonumber\\
	[L_{ud}^{S8,RR}]_{prst} &= V_{pi} [C_{quqd}^{(8)}]_{irst} \nonumber\\
	[L_{uddu}^{S1,RR}]_{prst} &= - V_{pi} [C_{quqd}^{(1)}]_{stir} \nonumber\\
	[L_{uddu}^{S8,RR}]_{prst} &= - V_{pi} [C_{quqd}^{(8)}]_{stir} \nonumber
\end{align}
The SMEFT coefficients must then be RG evolved from the EW scale up to the scale at which the heavy states are integrated out. To this aim we use DsixTools \cite{Celis:2017hod,Fuentes-Martin:2020zaz}.

As example we show the case of the diquark benchmark model, for which the only relevant operators are $[\mathcal{O}_{quqd}^{(1)}]$ and $[\mathcal{O}_{quqd}^{(8)}]$. The four coefficients contributing to the process from the matching in Eq.~\eqref{eq:SMEFT_tree_diquark} and only non-vanishing couplings $y^L_{12}$ $y^R_{13}$ are
\be
	\vec{C} = \left( [C_{quqd}^{(1)}]_{1123}, [C_{quqd}^{(8)}]_{1123}, [C_{quqd}^{(1)}]_{2113}, [C_{quqd}^{(8)}]_{2113} \right)^t~.
\ee
They evolve from $m_Z$ up to $m_{\Phi_6}$ as $\vec{C} (m_Z) = \mathcal{U}(m_Z, m_{\Phi_6}) \vec{C} (\Phi_6)$, with
\be
\hspace{-1cm}{\tiny 
\mathcal{U}(m_Z, M_{\Phi_6}) = 
\left(
\begin{array}{cccc}
 1.39 & -0.03 & 0.18 & 0.08 \\
 -0.2 & 0.92 & -0.18 & 0.08 \\
 0.18 & 0.08 & 1.39 & -0.03 \\
 -0.18 & 0.08 & -0.2 & 0.92 \\
\end{array}
\right)_{1 \TeV} ~, \quad
\left(
\begin{array}{cccc}
 1.67 & -0.04 & 0.33 & 0.12 \\
 -0.35 & 0.88 & -0.33 & 0.12 \\
 0.33 & 0.12 & 1.67 & -0.04 \\
 -0.33 & 0.12 & -0.35 & 0.88 \\
\end{array}
\right)_{5 \TeV}~
}
\ee
where we show the evolution for two values of the diquark mass.
In the case of the scalar doublet more operators are generated and the RG evolution has been performed using DsixTools.

\section{Details on the tree-level mediators}
\label{sec:mediators}

\subsection*{Conventions for dijet limits} \label{sec:defDJ}

Let us define here the couplings $x_{ij}$ of a bosonic resonance $X$ with quark bilinears used in the dijet analysis.
For a given representation of the resonance (spin, SU(3)$_c$) and a coupling to $q_i q^\prime_j$, with $q^{(\prime)} = u, d$ of arbitrary flavors $i$, $j$, we define the interaction Lagrangians as\footnote{Flavor matrices $x_{ij}$ are arbitrary complex matrices unless $q=q'$ when $x_{ij}$ is symmetric for scalar sextet, anti-symmetric for scalar triplet, and Hermitian for real vector singlet and octet where $+\text{h.c.}$ is removed from the lagrangian.}
\be\begin{split}
     (0, {\bf 1}): \qquad & \mathcal{L} \supset x_{ij} \, X \, \bar{q}_i P_X q^\prime_j \; + \text{h.c.}~, \\
     (0, {\bf 3}): \qquad & \mathcal{L} \supset x_{ij} \, X^\alpha \, \epsilon_{\alpha\beta\gamma} \, \bar{q^c}^{ \beta}_i P_X q^{\prime \gamma}_j + \text{h.c.}~, \\
     (0, {\bf 6}): \qquad & \mathcal{L} \supset x_{ij} \, X^m \,  S^m_{\alpha\beta} \, \bar{q^c}^{(\alpha|}_i P_X q^{\prime |\beta)}_j + \text{h.c.}~, \\
     (0, {\bf 8}): \qquad & \mathcal{L} \supset x_{ij} \, X^A \,  \bar{q}_i T^A P_X q^\prime_j \; + \text{h.c.}~, \\
     (1, {\bf 1}): \qquad & \mathcal{L} \supset x_{ij} \, X_\mu \,  \bar{q}_i \gamma^\mu P_X q^\prime_j \; (+ \text{h.c.})~, \\
     (1, {\bf 3}): \qquad & \mathcal{L} \supset x_{ij} \, X_\mu^\alpha \,  \epsilon_{\alpha\beta\gamma} \, \bar{q^c}^{\beta}_i \gamma^\mu P_X q^{\prime \gamma}_j + \text{h.c.}~, \\
     (1, {\bf 6}): \qquad & \mathcal{L} \supset x_{ij} \, X_\mu^m \,  S^m_{\alpha\beta} \, \bar{q^c}^{(\alpha|}_i \gamma^\mu P_X q^{\prime |\beta)}_j + \text{h.c.}~, \\
     (1, {\bf 8}): \qquad & \mathcal{L} \supset x_{ij} \, X_\mu^A \,  \bar{q}_i T^A \gamma^\mu P_X q^\prime_j \; (+ \text{h.c.})~, 
     \label{eq:X_lagrangians}
\end{split}\ee
where the chirality projector $P_X$ can be either $P_{L/R}$ for left/right spinors. Also, $T^A$ are the generators of $SU(3)_c$, $\psi_i^{(\alpha|} \psi_j^{|\beta)} = \frac{1}{2} (\psi_i^{\alpha} \psi_j^{\beta} + \psi_i^{\beta} \psi_j^{\alpha})$, and $S^m_{\alpha\beta}$ are the symmetric color matrices
\be\begin{split}
&	S^1 = \left( \begin{array}{ccc}
		1 & 0 & 0 \\
		0 & 0 & 0 \\
		0 & 0 & 0 
	\end{array} \right)~, \qquad
	S^2 = \frac{1}{\sqrt{2}} \left( \begin{array}{ccc}
		0 & 1 & 0 \\
		1 & 0 & 0 \\
		0 & 0 & 0 
	\end{array} \right)~, \qquad
	S^3 = \left( \begin{array}{ccc}
		0 & 0 & 0 \\
		0 & 1 & 0 \\
		0 & 0 & 0 
	\end{array} \right)~, \\ 
&	S^4 =  \frac{1}{\sqrt{2}}  \left( \begin{array}{ccc}
		0 & 0 & 0 \\
		0 & 0 & 1 \\
		0 & 1 & 0 
	\end{array} \right)~, \qquad
	S^5 = \left( \begin{array}{ccc}
		0 & 0 & 0 \\
		0 & 0 & 0 \\
		0 & 0 & 1 
	\end{array} \right)~, \qquad
	S^6 =  \frac{1}{\sqrt{2}} \left( \begin{array}{ccc}
		0 & 0 & 1 \\
		0 & 0 & 0 \\
		1 & 0 & 0 
	\end{array} \right)~,
\end{split}\label{eq:SymmColMatr}\ee
which satisfy
the matrices satisfy
\be
	\text{Tr} S^m \bar S_n = \delta^m_{\, n}~, \qquad \sum_m S^m_{\alpha\beta} \bar S^{\gamma\delta}_m = \frac{1}{2} (\delta_\alpha^\delta \delta_\beta^\gamma + \delta_\alpha^\gamma \delta_\beta^\delta )~.
\ee
where the conjugate matrices are given by $\bar S^{\alpha\beta}_m = S^m_{\alpha\beta}$.

\subsection*{Scalar color-sextet $\Phi_6 = (6, 1, 1/3)$}
\label{sec:sextet}

Let us start the study of the tree-level mediators listed in Eq.~\eqref{eq:mediators} with the singlet sextet diquark $\Phi_6 = ({\bf 6}, {\bf 1}, 1/3)$. This state has also been studied in \cite{Han:2009ya,Chen:2018stt}.
The relevant interaction Lagragian is:
\be
	\LL_{} \supset y^L_{ij}  {\bf \Phi}_6^{\alpha\beta \dagger} \bar{q}^{c, (\alpha|}_{L i} (i \sigma_2) q^{|\beta)}_{Lj} + y^R_{ij} {\bf \Phi}_6^{\alpha\beta \dagger} \bar{u}^{c (\alpha|}_{R i} d^{|\beta)}_{Rj} + {\rm h.c.}~,
\ee
where $y^L$ is an antisymmetric matrix. The components of the sextet representations are given as ${\bf \Phi}_6^{\alpha\beta} \equiv S^{m}_{\alpha\beta} \Phi_6^m$, where $m = 1, \ldots, 6$ and the symmetric color matrices $S^m_{\alpha\beta}$ are given in Eq.~\eqref{eq:SymmColMatr}.
The conjugate representation is given by ${\bf \Phi}_6^{\alpha\beta \dagger} = \bar{S}^{\alpha\beta}_m \Phi_6^{m *} = S_{\alpha\beta}^m \Phi_6^m$.
Matching $\Phi_6$ to the SMEFT at tree-level one has \cite{deBlas:2017xtg}:
\be\begin{split}
	[C_{qq}^{(1)}]_{ijkl} &= [C_{qq}^{(3)}]_{ilkj} = \frac{y^{L *}_{ik} y^{L}_{jl}}{4 M_{\Phi_6}^2}~, \\
	[C_{ud}^{(1)}]_{ijkl} &= \frac{2}{3} [C_{ud}^{(8)}]_{ijkl} = \frac{y^{R}_{jl} y^{R *}_{ik}}{3 M_{\Phi_6}^2}~, \\
	[C_{quqd}^{(1)}]_{ijkl} &= \frac{2}{3} [C_{quqd}^{(8)}]_{ijkl} = 4 \frac{y^L_{ki} y^{R *}_{jl}}{3 M_{\Phi_6}^2}~.
	\label{eq:SMEFT_tree_diquark}
\end{split}\ee
In terms of the $a$ EFT coefficients at the $m_b$ scale:
\be\begin{split}
	a_{V_{LL}}^{cb\alpha u} &= - \frac{4}{3} \kappa_{\rm RGE}^{V} \sum_{i \neq \alpha; \, j=1,2} V_{ci} V^*_{u j} \frac{ (y^{L *}_{\alpha i} y^{L }_{j 3})}{M_{\Phi_6}^2} ~, \\
	a_{V_{RR}}^{cb\alpha u} &= \frac{1}{3} \kappa_{\rm RGE}^{V} \frac{ (y^{R *}_{2 \alpha} y^{R }_{1 3})}{M_{\Phi_6}^2} ~, \\
	a_{S_{RR}}^{cb\alpha u} &= \frac{2}{3} \kappa_{\rm RGE}^{S} \sum_{i \neq \alpha} V_{c i} \frac{  y^{L *}_{\alpha i} y^{R}_{13}}{ M_{\Phi_6}^2}~, \\
	a_{S_{RR}}^{u\alpha b c} &= - \frac{2}{3} \kappa_{\rm RGE}^{S} \sum_{i = 2,3} V_{u i} \frac{ y^{L *}_{i 3} y^{R}_{2 \alpha} }{ M_{\Phi_6}^2}~, 
	\label{eq:aH4}
\end{split}\ee
where we already imposed that $y^{L}$ is antisymmetric and $\kappa_{\rm RGE}^{V,S}$ describe the effect of RGE from $m_{\Phi_6}$ to $m_b$. For instance $\kappa_{\rm RGE}^{S} \approx 1.65 \, (1.85) $ for $m_{\Phi_6} = 1 \, (5) \TeV$.
%

\subsection*{Scalar color-triplet $\Phi_3 = ({\bf \bar 3}, {\bf 1}, {\bf 1/3})$}

The interaction Lagragian to SM quarks is:\footnote{This state could also potentially couple to quarks and leptons, as a leptoquark \cite{deBlas:2017xtg}. Allowing for such couplings, together with the diquark ones, would induce proton decay. To avoid this we must thus assign baryon number $B(\Phi_3) = 2/3$ and impose at least $B$ conservation.}
\be
	\LL_{\Phi_3} \supset y^{qq}_{ij} \epsilon_{\alpha\beta\gamma} \Phi_3^\alpha \bar{q}^\beta_{Li} (i \sigma_2) q^{c\, \gamma}_{Lj} + y^{du}_{ij} \epsilon_{\alpha\beta\gamma} \Phi_3^\alpha \bar{d}^\beta_{Ri} u^{c\, \gamma}_{Rj} 
	+ {\rm h.c.} ~,
\ee
where $y^{qq}_{ij}$ is a symmetric matrix.
Matching to the $\Phi_3$ to the SMEFT at tree-level one has \cite{deBlas:2017xtg}:
\be\begin{split}
	[C_{qq}^{(1)}]_{ijkl} &= - [C_{qq}^{(3)}]_{ilkj} = \frac{y^{qq}_{ik} y^{qq *}_{lj}}{2 M_{\Phi_3}^2}~, \\
	[C_{ud}^{(1)}]_{ijkl} &= - \frac{1}{3} [C_{ud}^{(8)}]_{ijkl} = \frac{y^{du}_{ki} y^{du *}_{lj}}{3 M_{\Phi_3}^2}~, \\
	[C_{quqd}^{(1)}]_{ijkl} &= - \frac{1}{3} [C_{quqd}^{(8)}]_{ijkl} = 4 \frac{y^{qq}_{ki} y^{du *}_{lj}}{3 M_{\Phi_3}^2}~.
\end{split}\ee
No $\Delta F=2$  processes are generated at tree-level.
In terms of the $a$'s coefficients, one has (keeping into account the symmetricity of $y^{qq}$):
\be\begin{split}
	a_{V_{LL}}^{c b \alpha u} &= - \frac{4}{3} \kappa_{\rm RGE}^{V_{LL}} V_{c i} V^*_{u j} \frac{ y^{qq *}_{j3} y^{qq}_{i \alpha} }{ M_{\Phi_3}^2}~, \\
	a_{S_{RR}}^{c b \alpha u} &= - \frac{2}{3} \kappa_{\rm RGE}^{S} V_{c i} \frac{ y^{du *}_{31} y^{qq}_{i \alpha} }{ M_{\Phi_3}^2}~, \\
	a_{S_{LL}}^{c b \alpha u} &= - \frac{2}{3} \kappa_{\rm RGE}^{S} V_{u i} \frac{ y^{du *}_{\alpha 2} y^{qq}_{i 3} }{ M_{\Phi_3}^2}~, 
	\label{eq:aS1}
\end{split}\ee
where $ \kappa_{\rm RGE}^{V_{LL}} \approx 1.56$ and $\kappa_{\rm RGE}^{S} \approx 4.02$ for $M_{\Phi_3} = 1 \TeV$.
We consider two possible coupling structures to fit the anomaly.

{\bf 1)} Benchmark $V_{LL}$ -~
Setting $ y^{du} = 0$, we only generate the $a_{V_{LL}}$ coefficients. To minimise the impact of dijet bounds we must avoid a strong CKM suppression in the low-energy coefficients. This can be achieved with three non-vanishing couplings:
\be
	a_{V_{LL}}^{c b d u} = - 1.98 \frac{ y^{qq *}_{13} y^{qq}_{12} }{ M_{\Phi_3}^2} \approx \frac{0.23 V_{ud}}{\TeV^2}, \quad
	a_{V_{LL}}^{c b s u} = \frac{ - 1.98 y^{qq *}_{13} y^{qq}_{22} + 0.46 y^{qq *}_{13} y^{qq}_{21} }{ M_{\Phi_3}^2} \approx \frac{0.24 V_{us}}{\TeV^2}.
\ee
Note that the coupling to third generation $y^{qq}_{13}$ induces a decay of $\Phi_3$ to tops, which will put a constraint on the model even stronger than the dijet one.

{\bf 2)} Benchmark $S_{RR}$ -~
If the only non-zero RH coupling is $y^{du}_{31}$, then $a_{S_{LL}}^{cb\alpha u} = 0$ and we could get a good fit via the $a_{S_{RR}}$ coefficients. A strong CKM suppression is avoided with the couplings:
\be
	a_{S_{RR}}^{c b d u} = - 2.6 \frac{ y^{qq}_{12} y^{du *}_{31} }{ M_{\Phi_3}^2} \approx \frac{0.26 V_{ud}}{\TeV^2}, \quad
	a_{S_{RR}}^{c b s u} = \frac{ (- 2.6 y^{qq}_{22} + 0.60 y^{qq}_{12} ) y^{du *}_{31}}{ M_{\Phi_3}^2} \approx \frac{0.31 V_{us}}{\TeV^2}.
\ee
The best-fit for the anomalies is obtained for $y^{qq}_{22} \approx 0.50 y^{qq}_{12}$. In this scenario the coupling to the top quark is suppressed by $V_{ts}$, strongly reducing the relative branching ratio and thus the corresponding constraints. It is thus more favorable than the $V_{LL}$ benchmark.

\subsection*{Scalar color-triplet $\Psi_3 = ({\bf \bar 3}, {\bf 3}, {\bf 1/3})$}

This scalar couples to quarks as
\be
	\LL_{\Psi_3} \supset y^{qq}_{ij} \epsilon_{\alpha\beta\gamma} \Psi_3^{A\alpha} \bar{q}^\beta_{Li} \sigma^A (i \sigma_2) q^{c\, \gamma}_{Lj} 
	+ {\rm h.c.} ~,
\ee
where $y^{qq}_{ij}$ is an antisymmetric matrix.
The SMEFT coefficients generated integrating out this state at tree-level are \cite{deBlas:2017xtg}:
\be
	[C_{qq}^{(1)}]_{ijkl} = 3 [C_{qq}^{(3)}]_{ijkl} = 3\frac{y^{qq}_{ki} y^{qq *}_{lj}}{2 M_{\Psi_3}^2}~.
\ee
No $\Delta F=2$  processes are generated at tree-level, since $\Phi_3$ couples up to down quarks only.
A potentially good benchmark to fit the anomaly is with two non-vanishing couplings $y^{qq}_{12}$ and $y^{qq}_{23}$, giving:
\be
	a_{V_{LL}}^{c b d u} = \frac{4}{3} \kappa_{\rm RGE}^{V_{LL}} V_{us}^* V_{cs} \frac{ y^{qq *}_{23} y^{qq}_{12} }{ M_{\Psi_3}^2}~, \quad
	a_{V_{LL}}^{c b s u} = \frac{4}{3} \kappa_{\rm RGE}^{V_{LL}} V_{us}^*  \frac{ y^{qq *}_{23} ( - V_{cd} y^{qq}_{12} + V_{cb} y^{qq}_{23} )  }{ M_{\Psi_3}^2}~.
\ee
This setup is analogous to the $V_{LL}$ benchmark of $\Phi_3$, including a decay to top quarks induced by $y^{qq}_{23}$. They share a similar phenomenology, except for the fact that the weak-triplet has a stronger constraint from pair-production of dijet resonances, see \sec{sec:highpTpair}. For these reasons we do not consider this model further.

\subsection*{Scalar color-singlet $\Phi_1 = (1, 2, 1/2)$}

This scalar has the same quantum numbers as the SM Higgs. To avoid potentially very strong constraints while at the same time being able to fit the observed low-energy anomaly we must avoid that this state mixes too strongly with the SM Higgs and that it takes a non-zero vacuum expectation value. This can be achieved by appropriately tuning the scalar potential.

The Yukawa couplings to quarks are
\begin{equation}
	\mathcal{L}_{\Phi_1}^{\rm Yuk} = y^{qd}_{ij} \, \Phi_1^{\dagger}\bar d^i_R  q^j_L  \,  
	+ y^{qu}_{ij} \, \Phi_1^{\dagger} \epsilon \bar{q}^i_L u_R^j
	+ {\rm h.c.},
\end{equation}
Matching at tree-level to the SMEFT, the Yukawa couplings induce the following coefficients \cite{deBlas:2017xtg}:
\be\begin{split}
	[C_{quqd}^{(1)}]_{ijkl} &= -  \frac{  y^{ qu}_{ij} y^{dq *}_{lk}}{M_{\Phi_1}^2}~, \\
	[C_{qd}^{(1)}]_{ijkl} &= \frac{1}{6} [C_{qd}^{(8)}]_{ijkl} = - \frac{1}{6} \frac{ y^{ dq *}_{li} y^{ dq}_{kj}}{M_{\Phi_1}^2}~, \\
	[C_{qu}^{(1)}]_{ijkl} &= \frac{1}{6} [C_{qu}^{(8)}]_{ijkl} = - \frac{1}{6} \frac{ y^{ qu *}_{jk} y^{ qu}_{il}}{M_{\Phi_1}^2}~. \\
\end{split}\ee
While in general the neutral component can induce meson mixing at tree-level, this can be avoided by a suitable alignment of the couplings. The two benchmarks we consider, and the corresponding matching to the low-energy EFT, are discussed in Section~\ref{sec:Higgs}.

\subsection*{Scalar color-octet $\Phi_8 = (8, 2, 1/2)$}

This state has the same electroweak quantum numbers as the SM Higgs, but it is in the adjoint representation of $SU(3)_c$. It contains a neutral component that can potentially mediate meson mixing at tree-level. 
The interaction Lagrangian is
\be
	\LL_{\Phi_8} \supset y^{qu}_{ij} \Phi_8^{A \dagger} i \sigma_2 \bar{q}^T_{L i} T^A u_{R j} + y^{dq}_{ij} \Phi_8^{A \dagger} \bar{d}_{R i} T^A q_{L j}  + {\rm h.c.}~,
\ee
Matching to the SMEFT at tree-level one has \cite{deBlas:2017xtg}:
\be\begin{split}
	[C_{quqd}^{(8)}]_{ijkl} &= -  \frac{ y^{ dq *}_{lk} y^{ qu}_{ij}}{M_{\Phi_8}^2}~, \\
	[C_{qd}^{(1)}]_{ijkl} &= - \frac{4}{3} [C_{qd}^{(8)}]_{ijkl} = - \frac{2}{9} \frac{ y^{ dq *}_{li} y^{ dq}_{kj}}{M_{\Phi_8}^2}~, \\
	[C_{qu}^{(1)}]_{ijkl} &= - \frac{4}{3} [C_{qu}^{(8)}]_{ijkl} = - \frac{2}{9} \frac{ y^{ qu *}_{jk} y^{ qu}_{il}}{M_{\Phi_8}^2}~. \\
\end{split}\ee
The low-energy EFT coefficients relevant for a successful fit to the anomaly are given by
\be
	a_{S_{RR}}^{c b \alpha u} = \frac{2}{9} \kappa_{\rm RGE}^{S}  \frac{ y^{ dq *}_{b \alpha} V_{c i}  y^{ qu}_{i u}}{ M_{\Phi_8}^2}~, 
\ee
where $\alpha = 1,2$ and $\kappa_{\rm RGE}^{S} = 1.98 (2.27)$ for a mass of 1 TeV (5 TeV).
To minimise CKM suppression and the effect of the dijet constraints one can consider the following non-vanishing couplings: $y^{ dq}_{31}$, $y^{ dq}_{32}$, and $y^{qu}_{21}$. With this choice there is no contribution to meson mixing at tree level.

\subsection*{Vectors $\QQ_3 = (3, 2, 1/6)$ and $\QQ_6 = (\bar 6, 2, 1/6)$}

The colored vectors $\QQ_3$ and $\QQ_6$ couple to SM quarks as
\be
	\LL_{\QQ} \supset g^{\QQ_3}_{ij}  \QQ_3^{\mu \dagger} \epsilon_{ABC} \bar{d}^{B}_{R i} \gamma_\mu (i \sigma_2) q^{c C)}_{Lj} 
	    + \frac{1}{2} g^{\QQ_6}_{ij}  \QQ_6^{AB \mu \dagger} \bar{d}^{(A|}_{R i} \gamma_\mu (i \sigma_2) q^{c |B)}_{Lj} + h.c.
\ee
Matching to the SMEFT at tree-level one has \cite{deBlas:2017xtg}, for $\QQ_3$
\be\begin{split}
[C_{qd}^{(1)}]_{ijkl} &= - \frac{1}{3} [C_{qd}^{(8)}]_{ijkl} = \frac{2}{3} \frac{g^{\QQ_3}_{ki} g^{\QQ_3 *}_{lj}  }{ M_{\QQ_3}^2}~, 
\end{split}\ee
and for the $\QQ_6$
\be\begin{split}
	[C_{qd}^{(1)}]_{ijkl} &= \frac{2}{3} [C_{qd}^{(8)}]_{ijkl} = \frac{2}{3} \frac{g^{\QQ_6}_{ki} g^{\QQ_6 *}_{lj}  }{ M_{\QQ_6}^2}~, 
\end{split}\ee
In terms of the low-energy EFT coefficients we have
\be
	a_{S_{RL}}^{c b\alpha u} = \frac{4}{3} \kappa_{\rm RGE}^{S} V^*_{u i} V_{cj} \left(\frac{ g^{\QQ_3 *}_{3i} g^{\QQ_3}_{\alpha j} }{ M_{\QQ_3}^2} - \frac{ g^{\QQ_6 *}_{3i} g^{\QQ_6}_{\alpha j} }{ M_{\QQ_6}^2} \right)~,
\ee
where $\kappa_{\rm RGE}^{S} \approx 2.23$ for a UV mass scale of 1 TeV. The two states give the same contribution, up to a change of sign for one coupling, so we can focus on $\mathcal{Q_3}$ since it has weaker bounds from QCD pair production.
The combination with less CKM suppression is obtained with these three couplings only: $g^{\QQ_3, dq}_{31}$, $g^{\QQ_3, dq}_{12}$, and $g^{\QQ_3, dq}_{22}$:
\be
	a_{S_{RL}}^{c b d u} = \frac{2.8 V_{cs} V_{ud}^*}{M_{\QQ_3}^2} g^{\QQ_3 *}_{31} g^{\QQ_3}_{12} \approx - \frac{0.26 V_{ud}}{\TeV^2}~, \quad
	a_{S_{RL}}^{c b s u} = \frac{2.8 V_{cs} V_{ud}^*}{M_{\QQ_3}^2} g^{\QQ_3 *}_{31} g^{\QQ_3}_{22}  \approx - \frac{0.31 V_{us}}{\TeV^2}~.
\ee
To fit the anomaly we take $g^{\QQ_3}_{22} = g^{\QQ_3}_{12} V_{us}^*/V_{ud}^*$.

\section{Flavor constraints on the scalar sextet $\Phi_6$}
\label{sec:SextetFlavor}

Let us now discuss possible constraints from low-energy processes on the solution to the anomaly obtained in the main text.

\subsection*{$\Delta F = 2$}

\begin{figure}[t]
\centering
\includegraphics[width=7cm]{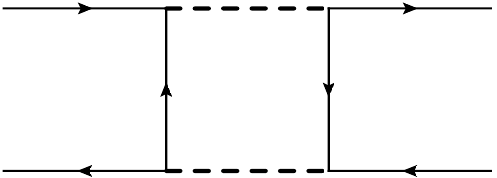} \quad
\includegraphics[width=7cm]{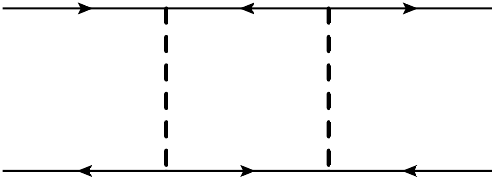} \\[0.5cm]
\includegraphics[width=7cm]{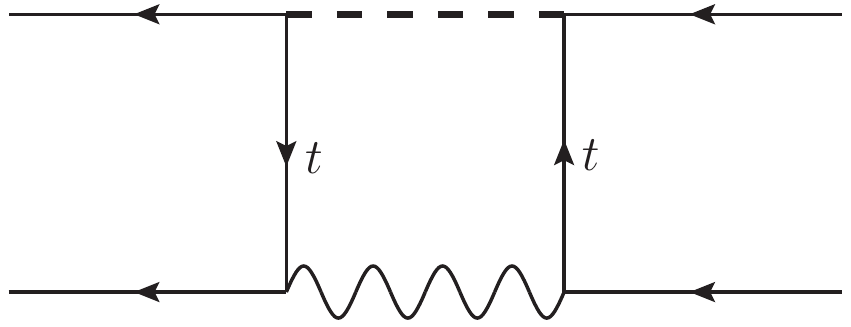}
\caption{\label{fig:boxesDF2} Diagrams inducing $\Delta F = 2$ processes in the sextet diquark model. The dashed line represents the $\Phi_6$ propagator.}
\end{figure}

Since the electric charge of $\Phi_6$ is $1/3$, no $\Delta F=2$  processes are generated at tree-level.
However, meson mixing can arise at the loop level via the diagrams shown schematically in Fig.~\ref{fig:boxesDF2}.
The boxes with two $\Phi_6$ propagators are proportional to the structure $(y^{L} y^{L \dagger})$ or $(y^{R} y^{R \dagger})$.
Given the effective Hamiltonian for down-quark $\Delta F=2$ processes, 
\be\begin{split}
	\mathcal{H}_{\Delta F=2} \supset& C_{VLL}^{q_i q_j} \left( \bar{q}^i_L \gamma_\mu q^j_L \right)^2 + C_{VRR}^{q_i q_j} \left( \bar{q}^i_R \gamma_\mu q^j_R \right)^2 + \\
		& + C_{LR1}^{q_i q_j} \left(\bar{q}^{i }_L \gamma_\mu q^{j}_L\right) \left(\bar{q}^{i }_R \gamma^\mu q^{j  }_R\right) +
		C_{LR2}^{q_i q_j} \left(\bar{q}^{i }_R q^{j}_L\right) \left(\bar{q}^{i }_L q^{j  }_R\right) ,
		\label{eq:hamiltonian_mixing}
\end{split}\ee
where $q = u,d$ and color is always contracted within the same current. 
Integrating out the diquark at one-loop level at the UV mass scale one has \cite{Chen:2018stt}:
\be\begin{split}
	C_{VLL}^{d_i d_j} &\approx \frac{3 (y_L^\dagger y_L)_{ij}^2}{16 \pi^2 M_{\Phi_6}^2}  - \frac{ (V y_L^*)_{3i} (V^* y_L)_{3j} (V^*_{ti} V_{tj}) }{8 \pi^2 M_{\Phi_6}^2} \left( 2 \frac{m_W^2}{v^2} 0.56 + \frac{m_t^4}{v^4} 1.48 \right), \\
	C_{VRR}^{d_i d_j} &= \frac{3 (y_R^\dagger y_R)_{ij}^2}{256 \pi^2 M_{\Phi_6}^2}, \\
	C_{LR1}^{d_i d_j} &\approx \frac{  (y_L^\dagger y_L)_{ij} (y_R^\dagger y_R)_{ij}}{64 \pi^2 M_{\Phi_6}^2} -  \frac{(y_R^\dagger y_R)_{ij} (V^*_{ti} V_{tj}) }{32 \pi^2 M_{\Phi_6}^2} \left( \frac{m_W^2 m_t^2}{v^4} 1.48 + \frac{m_t^2}{v^2 } 1.67 \right), \\
	C_{LR2}^{d_i d_j} &\approx  - \frac{10  (y_L^\dagger y_L)_{ij} (y_R^\dagger y_R)_{ij}}{64 \pi^2 M_{\Phi_6}^2} +  \frac{(y_R^\dagger y_R)_{ij} (V^*_{ti} V_{tj}) }{32 \pi^2 M_{\Phi_6}^2} \left( \frac{m_W^2 m_t^2}{v^4} 1.48 + \frac{m_t^2}{v^2 } 1.67 \right), 
\end{split}\ee
\be\begin{split}
	C_{VLL}^{u_i u_j} &\approx \frac{3 (V y_L^* y_L^T V^\dagger)_{ij}^2}{16 \pi^2 M_{\Phi_6}^2}  , \\
	C_{VRR}^{u_i u_j} &= \frac{3 (y_R^* y_R^T)_{ij}^2}{256 \pi^2 M_{\Phi_6}^2}, \\
	C_{LR1}^{u_i u_j} &\approx \frac{  (V y_L^* y_L^T V^\dagger)_{ij} (y_R^* y_R^T)_{ij}}{64 \pi^2 M_{\Phi_6}^2} , \\
	C_{LR2}^{u_i u_j} &\approx  - \frac{10  (V y_L^* y_L^T V^\dagger)_{ij} (y_R^* y_R^T)_{ij}}{64 \pi^2 M_{\Phi_6}^2}~. 
\end{split}\ee
Concerning the Wilson coefficient $C_{VLL}^{d_i d_j}$, the loop functions for the mixed $W-\Phi_6$ contributions have been computed explicitly, yielding
\begin{equation}
\begin{aligned}
C_{VLL}^{d_i d_j}\bigg|_{W-\Phi_6}=& -(N_c-1)  \frac{ (V y_L^*)_{3i} (V^* y_L)_{3j} (V^*_{ti} V_{tj}) }{16 \pi^2 }\frac{m_W^2}{v^2}m_t^2 \times\\
&\bigg[\frac{1}{(m_t^2-M^2)(m_t^2-m_W^2)}+\frac{m_t^2}{(m_t^2-M^2)^2(M^2-m_W^2)}\log\left(\frac{M^2}{m_t^2}\right) \\
&+\frac{m_t^2}{(m_t^2-m_W^2)^2(M^2-m_W^2)}\log\left(\frac{m_t^2}{m_W^2}\right)\bigg]
\end{aligned}
\end{equation}
The effect of the RGE from the diquark to the electroweak scale is mainly due to the QCD anomalous dimension. The $C_{VLL}^{q_i q_j}$ operators are rescaled by the factor $\eta^{6/21}$, where $\eta = \alpha_s(M_{\Phi_6}) / \alpha_s(m_t)$. The other anomalous dimensions are well known and can be seen also in \cite{Chen:2018stt}.

Particularly dangerous for a possible solution to the anomaly are the contributions from the left-handed coupling $y_L$. A way to avoid them is to require that  $(y^{L} y^{L \dagger}) = {\rm diag}(x,x,y)$, which is indeed realised by the coupling structure fixed in Eq.~\eqref{eq:S6Coupl}, which gives $(y^{L} y^{L \dagger}) =  {\rm diag}(|y^L_{12}|^2, |y^L_{12}|^2, 0)$.

\begin{figure}[t]
\centering
\includegraphics[height=6cm]{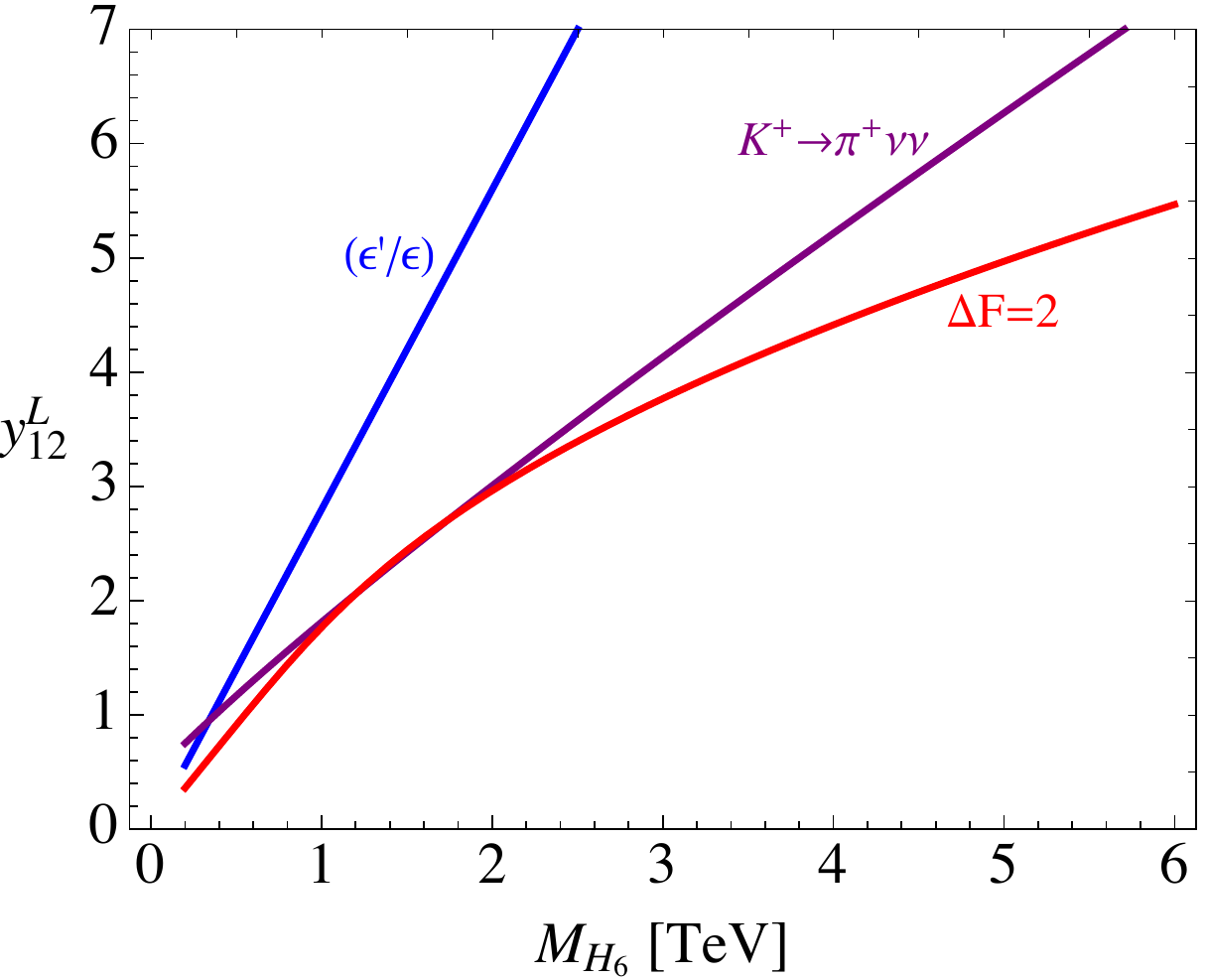} 
\caption{\label{fig:limitDF2} 95\% CL limit on $y^L_{12}$ from meson mixing (red), $K^+ \to \pi^+ \nu \nu$ (purple), and $(\epsilon^\prime / \epsilon)$ (blue), as a function of the diquark mass.}
\end{figure}

With this choice, the two diquark boxes in the top row of Fig.~\ref{fig:boxesDF2} give zero contribution to Kaon and $B_{d/s}$-mixing, while a strongly CKM-suppressed contribution to $D$-mixing.
The box with the $W$ boson shown in the bottom row of Fig.~\ref{fig:boxesDF2} gives a non-vanishing contribution only to Kaon mixing, which is also strongly CKM suppressed:
\be\begin{split}
	C_{VLL}^{sd} [M_{\Phi_6} ]  &= \frac{ |y^{L}_{12}|^2 (V^*_{ts} V_{td})^2 }{8 \pi^2 M_{\Phi_6}^2} \left( 2 \frac{m_W^2}{v^2} 0.56 + \frac{m_t^4}{v^4} 1.48 \right), \\
	C_{VLL}^{cu} [M_{\Phi_6} ] &= - \frac{3 |y^{L}_{12}|^4 (V_{cb} V^*_{ub})^2}{16 \pi^2 M_{\Phi_6}^2}  . \\
\end{split}\ee
Taking the low-energy matrix elements from \cite{Aebischer:2020dsw} and the constraints from UTFit \cite{Bona:2007vi},\footnote{We use the updated results shown by L. Silvestrini at the La Thuile conference in 2018.} we get the limit on $y^L_{12}$ as a function of the diquark mass as shown as a red line in Fig~\ref{fig:limitDF2}. 
This bound is reported as an upper limit on the product $y^{L*}_{12} y^R_{13}$ as a function of $m_{\Phi_6}$ of Fig.~\ref{fig:colored_vs_dijet} by setting $y^R_{13}$ to either its maximal possible value from perturbativity or to be equal to $y^L_{12}$.

\subsection*{$K^+ \to \pi^+ \nu \nu$}

The golden-channel Kaon decay can be described by the effective Hamiltonian \cite{Buras:1998raa}
\be
	\mathcal{H}_{\rm eff} \supset \frac{G_F V^*_{ts} V_{td}}{\sqrt{2}} \frac{\alpha}{\pi} C_{\nu_\alpha}^L (\bar{s}_L \gamma_\mu d_L) (\bar \nu_\alpha \gamma^\mu (1-\gamma_5) \nu_\alpha) + h.c.~,
\ee
where $\alpha = 1,2,3$ is the neutrino flavor and we do not consider lepton flavor violation.
The SM contribution is given by
\be
	C_{\nu_\alpha}^{L, SM} = - \frac{1}{s_W^2} \left( X_t + \frac{V_{cs}^* V_{cd}}{V^*_{ts} V_{td}} X_c^\alpha \right)~,
\ee
where $X_t \approx 1.48$, $X_c^e = X_c^\mu \approx 1.053 \times 10^{-3}$, and $X_c^\tau \approx 0.711 \times 10^{-3}$. 
The diquark contribution is lepton flavor universal and is given by \cite{Chen:2018stt}
\be
	C_{\nu}^{L, \Phi_6} = \frac{1}{2 s_W^2} \frac{v^2 (V y_L^*)_{32} (V^* y_L)_{31} }{m_W^2 V^*_{ts} V_{td} } I_Z\left(\frac{m_t^2}{M_{\Phi_6}^2}\right)~,
\ee
where the loop function is
\be
	I_Z(y_t) = - \frac{y_t}{1-y_t} - \frac{y_t \log y_t}{(1-y_t)^2}~.
\ee
The branching ratio is given by
\be
	\text{Br}(K^+ \to \pi^+ \nu \nu) = 2 \text{Br}(K^+ \to \pi^+ \nu_e \nu_e)_{\rm SM} \left| 1 + \frac{C_{\nu}^{L, \Phi_6}}{C_{\nu_e}^{L, SM} } \right|^2 + \text{Br}(K^+ \to \pi^+ \nu_\tau \nu_\tau)_{\rm SM} \left| 1 + \frac{C_{\nu}^{L, \Phi_6}}{C_{\nu_\tau}^{L, SM} } \right|^2~,
\ee
where $\text{Br}(K^+ \to \pi^+ \nu_e \nu_e)_{\rm SM} = 3.06 \times 10^{-11}$ and $\text{Br}(K^+ \to \pi^+ \nu_\tau \nu_\tau)_{\rm SM} = 2.52 \times 10^{-11}$.
The most recent measurement from NA62 is \cite{CortinaGil:2020vlo}\footnote{NA62 results on $K^+ \to \pi^+ \nu\nu$ have been updated in a \href{https://indico.cern.ch/event/868940/contributions/3815641/attachments/2080353/3496097/RadoslavMarchevski_ICHEP_2020.pdf}{presentation} at the ICHEP2020 conference.}
\be
	\text{Br}(K^+ \to \pi^+ \nu \nu) = (11.0~^{+ 4.0}_{- 3.5}) \times 10^{-11}
\ee
The limit on $y_{12}^L$ as a function of the diquark mass is shown as a purple line in Fig.~\ref{fig:limitDF2}.

\subsection*{$\epsilon^\prime / \epsilon$}

The decay amplitude $s \to d u \bar u$ is induced at tree-level in our setup, the imaginary part is strongly constrained by the direct CP violation effects in $\epsilon^\prime / \epsilon$.
A general master formula of $\epsilon^\prime / \epsilon$ in terms of EFT coefficients evaluated at the EW scale has been obtained in Ref.~\cite{Aebischer:2018quc}.
With the active couplings $y^{L}_{12}$ and $y^R_{13}$, the relevant operators are the purely left-handed vector-vector ones. In the notation of \cite{Aebischer:2018quc} one has
\be
	(\epsilon^\prime / \epsilon)_{\rm BSM} = \sum_{u_i=u,c} \left( P^{u_i}_{VLL} \Im \left[ C^{u_i}_{VLL} \right] + \widetilde P^q_{VLL} \Im \left[ \widetilde C^q_{VLL} \right] \right)~,
\ee
where the numerical coefficients are $P^{u}_{VLL} \approx -4.3$, $\widetilde P^{u}_{VLL} \approx 1.5$, $P^{c}_{VLL} \approx 0.7$, and $\widetilde P^{c}_{VLL} \approx 0.2$.
The rotation to the LEFT operator basis in Eq.~\eqref{eq:LEFT_basis} is given by
\be\begin{split}
	\mathcal{N} C^{u_i}_{VLL} &= [L^{V1LL}_{ud}]_{i i 2 1} - \frac{1}{6} [L^{V8LL}_{ud}]_{i i 2 1}~, \\
	\mathcal{N} \widetilde C^{u_i}_{VLL} &= \frac{1}{2} [L^{V8LL}_{ud}]_{i i 2 1}~,
\end{split}\ee
where $\mathcal{N} = (1 \TeV)^{-2}$. Matching to the SMEFT, Eq.~\eqref{eq:LEFtoSMEFT}, and then to the diquark model using Eq.~\eqref{eq:SMEFT_tree_diquark} and keeping only $y^L_{12}$ we get:
\be
	(\epsilon^\prime / \epsilon)_{\rm BSM} 
	\approx 1.3 \times 10^{-4} \frac{|y^L_{12}|^2 }{M_{\Phi_6}^2 / \TeV^2}~,
\ee
where the phase is only due to the CKM. Using the approximate upper bound for the BSM contribution of $(\epsilon^\prime / \epsilon)_{\rm BSM} \lesssim 10 \times 10^{-4}$ we get the constraint shown in Fig.~\ref{fig:limitDF2} as a blue line, which is weaker than those discussed above.

\section{Flavor constraints on the colorless scalar $\Phi_1$}
\label{sec:HiggsFlavor}
We discuss possible constraints from low-energy processes for the $\Phi_1$. In particular, we focus on a specific benchmark point for the solution of the anomaly: $M_{\Phi_1}\sim m_t$, $y^d_3 \sim 0.6$ and $y^d_2 = y^d_1 \sim 0.17$.\\ 
\paragraph{$\Delta F =2$}  The scalar $\Phi_1$ generates contributions to neutral meson mixing  at loop level through box diagrams. In the notation of \eq{eq:hamiltonian_mixing}, only the Wilson coefficients $C_{VRR}^{q_iq_j}$ and $C_{LR_2}^{q_iq_j}$ are non-zero.
Using the results in Refs.~\cite{Grossman:1994jb,Jung:2010ik} we get
\begin{align}
    C_{VRR}^{q_iq_j} =\,&  \frac{1}{128 \pi^2 M_{\Phi_1}^2} (V_{ti}V^*_{tj})^2 (y_j^{d*})^2(y_i^d)^2 y_t I_1(y_t) \,, \\
     C_{LR_2}^{q_iq_j} =\,&  \frac{1}{16\sqrt{2} \pi^2 M_{\Phi_1}^2} G_F m_W^2
     (V_{ti}V^*_{tj})^2 (y_j^{d*})(y_i^d) F(y_t,x_t) \,,
\end{align}
where $y_q = m_q^2/M_{\Phi_1}^2$, $x_q= m_q^2/m_{w}^2$. Using our benchmark point and the expression of the loop functions in Refs.~\cite{Grossman:1994jb,Jung:2010ik}, we have $I_1(1) = 1/3$ and $F(1,x_t) = 8.09$. We use the low-energy matrix elements from \cite{Aebischer:2020dsw}, finding that our results for this scenario are compatible with the current limits by UTFit \cite{Bona:2007vi} in both the $B_d$ and $B_s$ cases. Similar conclusions can be drawn in the case of neutral $K$ and $D$ mixings.

\paragraph{$Z\to b\bar b$} 
The contributions to $Z\to b\bar b$ decays come from penguin type diagrams. In particular, the $\Phi_1$ generates right-handed vector interactions, which yield
\begin{align}
    g_b^R \equiv&\, (g_b^R)_\text{SM}-\frac{(y_3^d)^2}{32\pi^2}\left[f_1(y_t)+\frac{\alpha_s}{3\pi}f_2(y_t)\right]\,.
\end{align}
The loop functions $f_1(y_t)$ and $f_2(y_t)$ are reported in Refs.~\cite{Jung:2010ik,Degrassi:2010ne} and in our benchmark point they assume the values $f_1(1)=1/2$ and $f_2(1)=-13/6$. We compare with the current value extracted by electroweak fits. In Ref.~\cite{ALEPH:2005ab}, the fitted value of $g_b^R = 0.0962\pm 0.0063$ is obtained. In this scenario, the contribution due to NP to $g_b^R$ is well below the uncertainty reported in Ref.~\cite{ALEPH:2005ab}.

\paragraph{$b\to s\ell^+\ell^-$} 
The contributions to $b\to s\ell^+\ell^-$ decays come from penguin type diagrams. We follow the conventions for the effective low-energy Hamiltonian in Ref.~\cite{Blake:2016olu}. The $\Phi_1$ generates the NP Wilson coefficients $\mathcal{C}_{7^\prime}$, $\mathcal{C}_{9^\prime}$ and $\mathcal{C}_{10^\prime}$, which are lepton-flavour universal. 
Concerning $\mathcal{C}_{9^\prime}$ and $\mathcal{C}_{10^\prime}$, using the results in \cite{Grinstein:1988me}, we get 
\begin{equation}
    \mathcal{C}_{9^\prime} (\mu_W)=-\mathcal{C}_{10^\prime}(\mu_W) = -\frac{1}{g^2 \sin^2\theta_w} y_2^d y_3^{d^*} B(y_t)\,,
\end{equation}
where $B(1)= -1/8$ and $\mu_W = m_W$. At the low scale $\mu_b = m_b$ we get
\begin{equation}
    \mathcal{C}_{9^\prime}(\mu_b)=-\mathcal{C}_{10^\prime}(\mu_b) \sim 0.13\,.
\end{equation}
The best sensitivity to these Wilson coefficients is achieved in $b\to s\mu^+\mu^-$ data, which show interesting deviations w.r.t. the SM expectations. The current status is found in \cite{Alguero:2019ptt,Aebischer:2019mlg,Ciuchini:2020gvn,Alok:2019ufo}, where several NP scenarios are analysed. Our predictions for $\mathcal{C}_{9^\prime}$ and $\mathcal{C}_{10^\prime}$ are not excluded, but are also not able to explain the tensions in $b\to s\mu^+\mu^-$ data.

The $\Phi_1$ generates also dipole operators and in particular $\mathcal{O}_{7^\prime}$ receives $m_b$ enhanced contributions. Following \cite{Besmer:2001cj}, we get
\begin{equation}
    \mathcal{C}_{7^\prime}(\mu_W) = -\frac{1}{2} y_2^d y_3^d \frac{v^2}{M_{\Phi_1}^2} \left[\frac{2}{3}F_1(y_t)+F_2(y_t) \right]\,,
\end{equation}
where $F_1(y_t) = F_2(y_t)\to 1/24$. The RGE evolution of $\mathcal{C}_{7^\prime}$ is the same as for $\mathcal{C}_{7}$. At the low scale $\mu_b = m_b$ and at leading order in QCD, we have \cite{Buchalla:1995vs}
\begin{equation}
   \frac{\mathcal{C}_{7^\prime}(\mu_b)}{\mathcal{C}_{7}(\mu_b)}  = 1.6 \%\,,
\end{equation}
which is below the current bound in Ref.~\cite{Aaij:2020umj}.

\bibliographystyle{JHEP}
\bibliography{paper}
	
\end{document}